\documentclass[ALICE,manyauthors]{cernphprep}
\usepackage[comma,square,numbers,sort&compress]{natbib}
\usepackage{hyperref}
\usepackage{lineno}
\usepackage{xspace}
\usepackage[dvipsnames]{xcolor}
\usepackage[T1]{fontenc}
\usepackage{orcidlink}

\begin{document}
%

\newcommand{\pp}           {pp\xspace}
\newcommand{\ppbar}        {\mbox{$\mathrm {p\overline{p}}$}\xspace}
\newcommand{\XeXe}         {\mbox{Xe--Xe}\xspace}
\newcommand{\PbPb}         {\mbox{Pb--Pb}\xspace}
\newcommand{\pA}           {\mbox{pA}\xspace}
\newcommand{\pPb}          {\mbox{p--Pb}\xspace}
\newcommand{\AuAu}         {\mbox{Au--Au}\xspace}
\newcommand{\dAu}          {\mbox{d--Au}\xspace}
\newcommand{\ccbar}        {\mbox{$\mathrm {c\overline{c}}$}\xspace}
\newcommand{\bbbar}        {\mbox{$\mathrm {b\overline{b}}$}\xspace}

\newcommand{\s}            {\ensuremath{\sqrt{s}}\xspace}
\newcommand{\snn}          {\ensuremath{\sqrt{s_{\mathrm{NN}}}}\xspace}
\newcommand{\pt}           {\ensuremath{p_{\rm T}}\xspace}
\newcommand{\meanpt}       {$\langle p_{\mathrm{T}}\rangle$\xspace}
\newcommand{\ycms}         {\ensuremath{y_{\rm CMS}}\xspace}
\newcommand{\ylab}         {\ensuremath{y_{\rm lab}}\xspace}
\newcommand{\etarange}[1]  {\mbox{$\left | \eta \right |~<~#1$}}
\newcommand{\yrange}[1]    {\mbox{$\left | y \right |~<~#1$}}
\newcommand{\dndy}         {\ensuremath{\mathrm{d}N_\mathrm{ch}/\mathrm{d}y}\xspace}
\newcommand{\dndeta}       {\ensuremath{\mathrm{d}N_\mathrm{ch}/\mathrm{d}\eta}\xspace}
\newcommand{\avdndeta}     {\ensuremath{\langle\dndeta\rangle}\xspace}
\newcommand{\dNdy}         {\ensuremath{\mathrm{d}N_\mathrm{ch}/\mathrm{d}y}\xspace}
\newcommand{\Npart}        {\ensuremath{N_\mathrm{part}}\xspace}
\newcommand{\Ncoll}        {\ensuremath{N_\mathrm{coll}}\xspace}
\newcommand{\dEdx}         {\ensuremath{\textrm{d}E/\textrm{d}x}\xspace}
\newcommand{\RpPb}         {\ensuremath{R_{\rm pPb}}\xspace}
\newcommand{\RPbPb}         {\ensuremath{R_{\rm PbPb}}\xspace}
\newcommand{\RAA}         {\ensuremath{R_{\rm AA}}\xspace}
\newcommand{\TAA}         {\ensuremath{T_{\rm AA}}\xspace}
\newcommand{\dsdy}         {\ensuremath{\textrm{d}\sigma/\textrm{d}y}\xspace}
\newcommand{\fb}           {\ensuremath{f_{\rm B}}\xspace}
\newcommand{\fbp}           {\ensuremath{f'_{\rm B}}\xspace}
\newcommand{\mee}          {\ensuremath{m_{\rm ee}}}
\newcommand{\x}            {\ensuremath{x}}
\newcommand{\fsig}         {\ensuremath{f_{\rm Sig}}}
\newcommand{\hb}           {\ensuremath{h_{\rm B}}}
\newcommand{\question}[1]     {\textcolor{blue}{#1}}
\newcommand{\update}[1]     {\textcolor{red}{#1}}

\newcommand{\nineH}        {$\sqrt{s}~=~0.9$~Te\kern-.1emV\xspace}
\newcommand{\seven}        {$\sqrt{s}~=~7$~Te\kern-.1emV\xspace}
\newcommand{\twoH}         {$\sqrt{s}~=~0.2$~Te\kern-.1emV\xspace}
\newcommand{\twosevensix}  {$\sqrt{s}~=~2.76$~Te\kern-.1emV\xspace}
\newcommand{\five}         {$\sqrt{s}~=~5.02$~Te\kern-.1emV\xspace}
\newcommand{\twosevensixnn}{$\sqrt{s_{\mathrm{NN}}}~=~2.76$~Te\kern-.1emV\xspace}
\newcommand{\fivenn}       {$\sqrt{s_{\mathrm{NN}}}~=~5.02$~Te\kern-.1emV\xspace}
\newcommand{\LT}           {L{\'e}vy-Tsallis\xspace}
\newcommand{\GeVc}         {Ge\kern-.1emV/$c$\xspace}
\newcommand{\MeVc}         {Me\kern-.1emV/$c$\xspace}
\newcommand{\TeV}          {Te\kern-.1emV\xspace}
\newcommand{\GeV}          {Ge\kern-.1emV\xspace}
\newcommand{\MeV}          {Me\kern-.1emV\xspace}
\newcommand{\GeVmass}      {Ge\kern-.2emV/$c^2$\xspace}
\newcommand{\MeVmass}      {Me\kern-.2emV/$c^2$\xspace}
\newcommand{\lumi}         {\ensuremath{\mathcal{L}}\xspace}

\newcommand{\ITS}          {\rm{ITS}\xspace}
\newcommand{\TOF}          {\rm{TOF}\xspace}
\newcommand{\ZDC}          {\rm{ZDC}\xspace}
\newcommand{\ZDCs}         {\rm{ZDCs}\xspace}
\newcommand{\ZNA}          {\rm{ZNA}\xspace}
\newcommand{\ZNC}          {\rm{ZNC}\xspace}
\newcommand{\SPD}          {\rm{SPD}\xspace}
\newcommand{\SDD}          {\rm{SDD}\xspace}
\newcommand{\SSD}          {\rm{SSD}\xspace}
\newcommand{\TPC}          {\rm{TPC}\xspace}
\newcommand{\TRD}          {\rm{TRD}\xspace}
\newcommand{\VZERO}        {\rm{V0}\xspace}
\newcommand{\VZEROA}       {\rm{V0A}\xspace}
\newcommand{\VZEROC}       {\rm{V0C}\xspace}
\newcommand{\Vdecay} 	   {\ensuremath{V^{0}}\xspace}

\newcommand{\ee}           {\ensuremath{{\rm e}^{+}{\rm e}^{-}}\xspace} 
\newcommand{\pip}          {\ensuremath{\pi^{+}}\xspace}
\newcommand{\pim}          {\ensuremath{\pi^{-}}\xspace}
\newcommand{\kap}          {\ensuremath{\rm{K}^{+}}\xspace}
\newcommand{\kam}          {\ensuremath{\rm{K}^{-}}\xspace}
\newcommand{\pbar}         {\ensuremath{\rm\overline{p}}\xspace}
\newcommand{\kzero}        {\ensuremath{{\rm K}^{0}_{\rm{S}}}\xspace}
\newcommand{\lmb}          {\ensuremath{\Lambda}\xspace}
\newcommand{\almb}         {\ensuremath{\overline{\Lambda}}\xspace}
\newcommand{\Om}           {\ensuremath{\Omega^-}\xspace}
\newcommand{\Mo}           {\ensuremath{\overline{\Omega}^+}\xspace}
\newcommand{\X}            {\ensuremath{\Xi^-}\xspace}
\newcommand{\Ix}           {\ensuremath{\overline{\Xi}^+}\xspace}
\newcommand{\Xis}          {\ensuremath{\Xi^{\pm}}\xspace}
\newcommand{\Oms}          {\ensuremath{\Omega^{\pm}}\xspace}
\newcommand{\degree}       {\ensuremath{^{\rm o}}\xspace}
\newcommand{\jpsi}        {\ensuremath{\rm J/\psi}\xspace}
\newcommand{\pJPsi}        {\ensuremath{\rm J/\psi}\xspace}
\newcommand{\DMeson}       {\ensuremath{\rm D}\xspace}

\begin{titlepage}
\PHyear{2023}       
\PHnumber{190}      
\PHdate{29 August}  

\title{Prompt and non-prompt \jpsi production at midrapidity in \PbPb collisions at \snn = 5.02 TeV}
\ShortTitle{Prompt and non-prompt \jpsi production in \PbPb collisions}   

\Collaboration{ALICE Collaboration\thanks{See Appendix~\ref{app:collab} for the list of collaboration members}}
\ShortAuthor{ALICE Collaboration} 

\begin{abstract}
	The transverse momentum (\pt) and centrality dependence of the nuclear modification factor \RAA of prompt and non-prompt \jpsi, the latter originating from the weak decays of beauty hadrons, have been measured by the ALICE collaboration in \PbPb collisions at \snn = 5.02 TeV. The measurements are carried out through the \ee decay channel at midrapidity ($|y| < 0.9$) in the transverse momentum region $1.5 < \pt < 10$~\GeVc. Both prompt and non-prompt \jpsi measurements indicate a significant suppression for \pt $>$ 5~\GeVc, which becomes stronger with increasing collision centrality. The results are consistent with similar LHC measurements in the overlapping \pt intervals, and cover the kinematic region down to \pt = 1.5 \GeVc at midrapidity, not accessible by other LHC experiments. The suppression of prompt \jpsi in central and semicentral collisions exhibits a decreasing trend towards lower transverse momentum, described within uncertainties by models implementing \jpsi production from recombination of c and $\overline{\rm c}$ quarks produced independently in different partonic scatterings. At high transverse momentum, transport models including quarkonium dissociation are able to describe the suppression for  prompt \jpsi. For non-prompt \jpsi, the suppression predicted by models including both collisional and radiative processes for the computation of the beauty-quark energy loss inside the quark--gluon plasma is consistent with measurements within uncertainties.    

\end{abstract}
\end{titlepage}

\setcounter{page}{2} 


\section{Introduction}
\label{Sec:Intro}

Quantum chromodynamics (QCD) calculations on the lattice~\cite{Karsch:2006xs,Borsanyi:2010bp,Borsanyi:2013bia,Bazavov:2011nk} predict the existence of the quark--gluon plasma (QGP), a state of strongly-interacting matter characterised by quark and gluon degrees of freedom, expected to be produced at extremely high temperatures and energy densities. Such conditions can be realised in the laboratory by colliding heavy ions at ultra-relativistic energies, enabling the study of the properties of this state of matter, as shown by multiple measurements carried out at the SPS~\cite{Csorgo:2000yu,Teaney:2000cw}, RHIC~\cite{BRAHMS:2004adc,PHOBOS:2004zne,STAR:2005gfr,PHENIX:2004vcz} and LHC~\cite{ALICE:2022wpn}. Heavy quarks, i.e.~charm and beauty, are mainly produced in hard parton--parton scatterings and on a shorter time scale than the QGP formation time at LHC energies ($\tau_{\rm QGP} \approx 1.5 {\rm fm}/c$)~\cite{Andronic:2015wma,Liu:2012ax}, thus experiencing the full QGP evolution. Charmonia, bound states of a charm and an anti-charm quark, are interesting probes of the QGP. In fact, it was predicted that their production would be suppressed in this medium due to static colour screening resulting from the high density of colour charges inside it~\cite{Matsui:1986dk} or due to dynamical dissociation~\cite{Rothkopf:2019ipj}. 
A suppression of the \jpsi yield in heavy-ion collisions relative to proton--proton collisions was observed at the SPS~\cite{BAGLIN1995617,NA50:2004sgj}, RHIC~\cite{PHENIX:2006gsi,PHENIX:2011img,STAR:2009irl,STAR:2013eve} and LHC~\cite{CMS:2012bms,ALICE:2012jsl,ALICE:2013osk,ALICE:2015jrl,ALICE:2016flj,ALICE:2019lga,ALICE:2019nrq,ATLAS:2018hqe,CMS:2017uuv,ALICE:2022inclJpsi}. However, at LHC energies, the \jpsi suppression is significantly reduced compared to lower energy results, in particular at low transverse momentum (\pt) and in more central collisions. These findings are interpreted by considering an additional contribution to the \jpsi production, known as regeneration, according to which the abundantly produced charm and anti-charm quarks from different hard partonic scatterings can recombine to form charmonium states~\cite{Braun-Munzinger:2000csl,PhysRevC.63.054905}. Previous ALICE inclusive \jpsi measurements at both central and forward rapidity, which have revealed significantly less suppression of \jpsi at low transverse momentum compared to lower energies~\cite{ALICE:2022inclJpsi,ALICE:2019lga,ALICE:2019nrq} as well as non-zero elliptic flow~\cite{ALICE:2020pvw,ALICE:2017quq}, have clearly demonstrated the relevance of this mechanism. 

Different phenomenological scenarios are assumed for the description of charmonium production in heavy-ion collisions. In the Statistical Hadronisation model~\cite{Braun-Munzinger:2000csl,Andronic:2019wva}, the abundances of all charmonium states are determined by thermal weights at the chemical freeze-out. By contrast, according to several partonic transport models~\cite{PhysRevC.63.054905,Zhou:2014kka,Du:2015wha} the charmonium states can be produced via regeneration throughout the full evolution of the QGP phase. Recent inclusive \jpsi results in \PbPb collisions at \snn = 5.02 TeV  indicate that both approaches can provide a description of the measured suppression within uncertainties in the low-\pt region~\cite{ALICE:2022inclJpsi,ALICE:2019lga,ALICE:2019nrq}. Results of $\psi(2S)$ suppression recently released by the ALICE collaboration~\cite{ALICE:2022jeh} show that transport models better reproduce charmonium measurements for central events compared to the Statistical Hadronisation model. Besides providing a good description of the production, transport models are also able to describe the elliptic flow of inclusive \jpsi~\cite{ALICE:2020pvw,He:2022ywp}. As the \pt of the \jpsi increases, the contribution from regeneration becomes less relevant, while charmonium dissociation and fragmentation of high-energy partons into charmonia become dominant.

Inclusive \jpsi production in high-energy hadronic collisions consists of several contributions: the \jpsi produced directly and from the decays of higher mass charmonium states (e.g. $\psi$(2S) or $\chi_{\rm c}$), known as the ``prompt'' contribution, and \jpsi originating from the weak decays of beauty hadrons. The latter component, referred to as ``non-prompt'', is characterised by a production vertex displaced with respect to the primary vertex of the collision, and this feature is exploited experimentally to separate the two contributions. The measurement of prompt \jpsi production enables a direct comparison with prompt charmonium models. 
In addition, the non-prompt \jpsi production measurement grants a direct insight into the suppression of beauty hadrons, 
which are expected to be sensitive to the properties of the medium created in heavy-ion collisions as ancestor beauty quarks experience energy loss by interacting with QGP constituents.
The energy loss of partons inside the medium is expected to happen via both radiative~\cite{Gyulassy:1990ye,Baier:1996sk} and elastic collisional processes~\cite{THOMA1991491,PhysRevD.44.1298,PhysRevD.44.R2625}. The relative contribution of the former is expected to increase with \pt. The energy loss strongly depends on the colour charge of the parton, being larger for gluons than for quarks, as well as on the parton mass~\cite{Dokshitzer:2001zm,Armesto:2003jh,Wicks:2007am}. The production of open heavy-flavour hadrons in nuclear collisions is also affected by in-medium hadronisation effects. Due to the high quark density in the QGP, heavy quarks can also hadronise via coalescence, by recombining with other light flavour quarks inside the medium~\cite{Fries:2003vb,Ravagli:2007xx,Greco:2003vf}. 

In order to interpret both prompt and non-prompt \jpsi results in nuclear collisions, cold nuclear matter (CNM) effects need to be considered. The main effect at LHC energies is represented by the modification of the parton distribution functions of protons and neutrons inside nuclei compared to the ones of the free nucleons~\cite{Armesto:2006ph}. These effects are usually addressed through measurements in proton--nucleus collisions at the same centre-of-mass energy and are expected to be effective below 3~\GeVc for prompt \jpsi, as shown by recent \jpsi results in p--Pb collisions at \snn = 5.02 TeV~\cite{ALICE:2021lmn}. For non-prompt \jpsi, a small suppression with no significant \pt dependence is observed, although with large uncertainties~\cite{ALICE:2021lmn}. 

At LHC energies in heavy-ion collisions, the production of prompt and non-prompt \jpsi was measured at midrapidity by the CMS~\cite{CMS:2012bms} and ALICE~\cite{ALICE:2015nvt} collaborations in \PbPb collisions at \snn = 2.76 TeV. At the centre-of-mass energy of \snn = 5.02 TeV in the central rapidity region, the ATLAS~\cite{ATLAS:2018hqe} and CMS~\cite{CMS:2017uuv} collaborations measured the suppression of prompt and non-prompt \jpsi at high transverse momentum. At low \pt, inclusive \jpsi measurements were carried out by the ALICE collaboration~\cite{ALICE:2022inclJpsi,ALICE:2019nrq}.
In this article, \pt and centrality dependent measurements 
of prompt and non-prompt \jpsi production at midrapidity in \PbPb collisions at \snn = 5.02 TeV are presented. These results, obtained down to \pt = 1.5 \GeVc, have a unique kinematic coverage at the LHC compared to existing midrapidity measurements at the same centre-of-mass energy, which are available only at high transverse momentum. Furthermore, they have a significantly improved precision compared to previous published ALICE results at lower energy. 

This article is organised as follows: the ALICE apparatus and data samples are described in Section~\ref{Sec:DataSample}, the analysis technique is presented in Section~\ref{Sec:DataAnalysis}, results and comparison with similar measurements from other experiments and model calculations are discussed in Section~\ref{Sec:discussion}, and finally the summary is provided in Section~\ref{Sec:summary}.

\section{Experimental apparatus and data sample}
\label{Sec:DataSample}

The ALICE detector consists of a central barrel with a pseudorapidity coverage of $|\eta| < 0.9$ and a forward rapidity muon spectrometer with a pseudorapidity coverage for muons of $-4 < \eta < -2.5$. It also includes forward and backward pseudorapidity detectors employed for triggering, background rejection, and event characterisation. Central barrel detectors are placed inside a magnetic field $B$ = 0.5 T provided by a solenoidal magnet. A complete description of the detector and an overview of its performance are discussed in Refs.~\cite{Collaboration_2008,ALICE:2014sbx}. 
The main detectors employed for the analysis described in this article are the Inner Tracking System (ITS)~\cite{ALICE:2014sbx}, the Time Projection Chamber (TPC)~\cite{ALME2010316} and the V0 detector~\cite{ALICE:2013axi}. Both ITS and TPC detectors 
enable the measurement of inclusive \jpsi mesons via the dielectron decay channel in the central rapidity region down to zero \pt. The ITS consists of six layers of silicon detectors, with
the two innermost layers composed of silicon pixel detectors (SPD) which provide the spatial resolution to separate prompt and non-prompt \jpsi on a statistical basis.
The TPC is the main tracking detector of the central barrel. In addition, it allows for particle identification via the measurement of the specific ionisation energy loss d$E$/d$x$ in the detector gas. 
The V0 detector consists of two scintillator arrays placed on each side of the interaction point (with pseudorapidity coverage $2.8 < \eta < 5.1$ and $-3.7 < \eta < -1.7$), 
and it is used to reject offline beam-induced background events, to define a minimum bias trigger, and to characterise the event centrality. 
The zero degree calorimeters~\cite{Cortese:2019nnv}, located at $\pm$112.5~m on both sides of the interaction point, are used to reject electromagnetic interactions and beam-induced background in Pb–Pb collisions.
The results presented in this article are based on the same data samples of \PbPb collisions at \snn = 5.02 TeV employed for the inclusive \jpsi analysis~\cite{ALICE:2022inclJpsi}. 
In particular, it consist of a combination of the data samples collected during the years 2015 and 2018 of the LHC Run~2. 
In order to obtain a uniform acceptance of the detectors, only events with a reconstructed primary vertex position
along the beam line located within $\pm$10 cm from the centre of the detector were considered. In addition, consecutive events triggered within a time interval smaller than the readout time of the TPC
were discarded as these are expected to have  a high total charge deposition in the active volume and consequently significantly worse particle identification performance of the TPC.
The datasets are collected with a minimum bias triggered event sample, provided by the coincidence of signals in the two scintillator arrays of the V0 detector.
The corresponding integrated luminosity is about 24 $\rm \mu b^{-1}$~\cite{ALICE-PUBLIC-2018-011}. In addition, centrality triggered samples, whose trigger definition is based on the amplitude of the signal collected
in the V0 detector, were used. These correspond to central (0–10\%) and semicentral (30–50\%) events, equivalent to an integrated luminosity of 105 and 51 $\rm \mu b^{-1}$~\cite{ALICE-PUBLIC-2018-011}, respectively.

\section{Data analysis}
\label{Sec:DataAnalysis}

In order to estimate hot nuclear matter effects, the nuclear modification factor can be defined as the production yield in \PbPb collisions at \snn = 5.02 TeV normalised to the reference production cross section in pp collisions at the same energy and scaled by the average nuclear overlap function $\langle \TAA \rangle$~\cite{ALICE-PUBLIC-2018-011}:

\begin{equation}
\label{eq:raa} 
	\RAA = \frac{{\rm d}N/{{\rm d}\pt{\rm d}{\it y}}}{\langle \TAA \rangle \times {\rm d}\sigma_{\rm pp}/{\rm d}\pt{\rm d}y}.
\end{equation}

Inclusive \jpsi measurements with no separation between prompt and non-prompt contributions are carried out by the ALICE collaboration at midrapidity  ($|y| < 0.9$) in \PbPb collisions at \snn = 5.02 TeV, as discussed in Ref.~\cite{ALICE:2022inclJpsi}. The nuclear modification factors of prompt and non-prompt \jpsi at midrapidity can be obtained by combining inclusive \jpsi \RAA measurements with non-prompt \jpsi fractions (\fb), the latter defined as the ratios of the production yields of \jpsi mesons originating from beauty-hadron decays to that of inclusive \jpsi, estimated in both \PbPb collisions and pp collisions at the same centre-of-mass energy.

\subsection{Non-prompt \jpsi analysis in \PbPb collisions}
\label{Sec:NonPromptAnalysisPbPb}

\subsubsection*{Selection of \jpsi candidates}

The event selection and track quality requirements used in the analysis discussed in this article are identical to those used for the corresponding midrapidity inclusive \jpsi \RAA analysis in \PbPb collisions at \snn = 5.02~TeV~\cite{ALICE:2022inclJpsi}. 
Prompt and non-prompt \jpsi measurements are carried out in the \jpsi transverse momentum interval 1.5--10~\GeVc and in four different centrality classes, namely 0--10\%, 10--30\%, 30--50\%, and 50--90\%. The results are presented as a function of the transverse momentum of the \jpsi, as well as of the average number of participants ($\langle N_{\rm part} \rangle$). The latter can
be estimated via Glauber model calculations~\cite{Glauber:1970jm,doi:10.1146/annurev.nucl.57.090506.123020,dEnterria:2020dwq,Loizides:2017ack} for the different centrality intervals~\cite{ALICE-PUBLIC-2018-011}.  

\jpsi candidates are reconstructed at midrapidity ($|y| < 0.9$) through the dielectron decay channel. Electron candidates, reconstructed using both ITS and TPC detectors, are required to have a pseudorapidity in the interval $|\eta| < 0.9$, a minimum transverse momentum of 1~\GeVc, and a minimum of 70 space points out of a maximum of 159 in the TPC.  A hit in at least one of the two SPD layers is also required to improve the tracking and the spatial resolution. Several quality selection criteria,  also employed for the inclusive \jpsi analysis~\cite{ALICE:2022inclJpsi}, are considered in order to ensure good track resolution, as well as to reduce the contribution of electrons and positrons originating from photon conversion in the detector material and to reject secondary tracks originating from weak decays.  
The electron identification relies on the measurement of d$\it E$/d$\it x$ in the TPC, using stricter requirements than those considered for the inclusive \jpsi analysis in order to increase the signal-to-background ratio, in particular at low transverse momentum and in central events. For tracks reconstructed in more peripheral centrality classes, namely 30--50\% and 50--90\%, the TPC d$\it E$/d$\it x$
signal is required to lie within the interval $[-3, 3]\sigma_{\rm e}$ relative to the expectation for electrons, where $\sigma_{\rm e}$ represents the specific energy-loss resolution for electrons in the TPC. In addition, tracks consistent with the pion and proton assumptions within 3.5$\sigma$ are rejected. For the centrality classes 0--10\% and 10--30\%, the electron inclusion is requested to be  $[-2, 3]\sigma_{\rm e}$ $ \left (  [-1, 3]\sigma_{\rm e} \right )$ for the \pt of the electron below (above) 5~\GeVc. A 4.5$\sigma$ rejection for protons and pions is applied only below 5~\GeVc of the electron momentum at the inner wall of the TPC, while above 5~\GeVc the rejection is not applied to avoid reductions in the electron reconstruction efficiency. \jpsi candidates are then obtained by considering all opposite-sign charged electron pairs. In 0–10\% most central collisions, only candidates where both electron decay tracks have a hit in the first layer of the SPD are considered to optimise both the spatial resolution and signal-to-background ratio. For other centrality classes as well as in the highest \pt interval (7--10~\GeVc) in the 0–10\% centrality class, where the signal-to-background ratio is better, the condition is released to increase the efficiency. 
 Pair candidates where neither of the decay products has a hit in the first layer of the SPD are excluded due to the poor resolution of the secondary vertex.

\subsubsection*{Separation of prompt and non-prompt \jpsi}

The measurement of the non-prompt  \jpsi fraction 
is obtained through an unbinned two-dimensional likelihood fit procedure applied to reconstructed \jpsi candidate pairs, following the same techniques employed in previous publications~\cite{ALICE:2015nvt,ALICE:2021lmn,ALICE:2022zig,ALICE:2018szk,ALICE:2021edd}. A simultaneous unbinned log-likelihood fit of the \jpsi candidate distribution as a function of invariant mass (\mee) and  pseudoproper decay length (\x) values is performed. The pseudoproper decay length is defined as $x = c \times L_{\rm xy} \times m_{\rm \jpsi}/\pt$, where $L_{\rm xy} =  (\Vec{L} \times \Vec{\pt})/\pt$ represents the projection in the transverse plane of the vector pointing from the primary vertex to the \jpsi decay vertex ($\Vec{L}$) and $m_{\rm \jpsi}$ is the \jpsi mass provided by the Particle Data Group (PDG)~\cite{Workman:2022ynf}. The fit procedure maximises the logarithm of a likelihood function

\begin{equation}
\label{eq:likeFunc} 
\ln{\mathcal{L}}= \sum_{i=1}^{N} \ln{\left[ \fsig \times F_{\rm Sig}(\x^{i}) \times M_{\rm Sig}(\mee^{i}) + (1 - \fsig) \times F_{\rm Bkg}(\x^{i}) \times M_{\rm Bkg}(\mee^{i})\right]},
\end{equation}

where $N$ is the number of \jpsi candidates within the invariant mass interval $2.72 < \mee < 3.40$ GeV/$c^{2}$ ($2.60 < \mee < 3.60$ GeV/$c^{2}$) in the centrality interval 0--10\% (10--90\%). A tighter invariant mass window is considered in most central collisions in order to increase the signal-to-background ratio in the fitting region, but still with a large enough sample of candidates to constrain the background probability density functions (PDFs). The relative amount of signal candidates, both prompt and non-prompt \jpsi, with respect to the total number of candidates is quantified by the fit parameter \fsig. The PDFs entering in Eq.~\ref{eq:likeFunc}, namely $F_{\rm Sig}(x)$ and $F_{\rm Bkg}(x)$ ($M_{\rm Sig}(x)$ and $M_{\rm Bkg}(x)$), are used to describe the pseudoproper decay length (invariant mass) distributions of signal and background, respectively. The pseudoproper decay length PDF of the signal is defined as

\begin{equation}
\label{eq:signalPseudoPropDL} 
F_{\rm Sig}(x) =  \fbp \times F_{\rm B}(x) + (1-\fbp) \times F_{\rm prompt}(x),
\end{equation}

where $F_{\rm B}(x)$ and $F_{\rm prompt}(x)$ are the $x$ PDFs for non-prompt and prompt \jpsi, respectively, while \fbp\ represents the fraction of \jpsi originating from beauty-hadron decays 
 in the sample of selected dielectron candidates. Both \fbp and \fsig\ are left as free parameters in the fitting procedure.
A correction to the \fbp fraction obtained from the fit is applied to take into account the different average acceptance-times-efficiencies of prompt and non-prompt \jpsi, which is a consequence of two effects: {\it (i)} different \jpsi \pt distributions inside the wide \pt intervals where the measurements are provided; {\it (ii) } different \jpsi polarisation, which can modify angular distributions of decay products and thus the \jpsi acceptance. The corrected fraction of non-prompt \jpsi, \fb, is obtained as

\begin{equation}
\label{eq:fBCorrProcedure} 
\fb = \left( 1 + \frac{1-\fbp}{\fbp} \times \frac{\langle A \times \epsilon \rangle_{\rm B~~~~~~}}{\langle A \times \epsilon \rangle_{\rm prompt}} \right)^{-1},
\end{equation}

where $\langle A \times \epsilon \rangle_{\rm prompt}$ and $\langle A \times \epsilon \rangle_{\rm B}$ represent the average acceptance-times-efficiency values for prompt and non-prompt \jpsi, respectively, in the considered \pt\ interval. The functional forms of the different PDFs in Eq.~\ref{eq:likeFunc} are determined either based on data or on Monte Carlo (MC) simulations, and are computed using the same procedures as in previous analyses~\cite{ALICE:2015nvt,ALICE:2021lmn,ALICE:2022zig,ALICE:2018szk}. The PDFs corresponding to the signal component, namely $F_{\rm prompt}(x)$, $F_{\rm B}(x)$, and $M_{\rm Sig}(\mee)$, as well as acceptance-times-efficiency corrections of prompt and non-prompt \jpsi in Eq.~\ref{eq:fBCorrProcedure}, are determined from MC simulations. These simulations consist of prompt and non-prompt \jpsi meson signals embedded in a background sample of \PbPb collisions at \snn = 5.02 TeV produced with HIJING~\cite{PhysRevD.44.3501}. The intervals of centrality for the \PbPb collisions considered in MC simulations are the same as those selected in experimental data. The prompt \jpsi component is simulated with a \pt spectrum based on existing inclusive \jpsi \PbPb measurements at midrapidity, while PYTHIA 6.4~\cite{Sjostrand:2006za} is used to generate beauty hadrons for non-prompt \jpsi simulations. The decays of beauty hadrons into final states containing a \jpsi meson are handled by the EvtGen R01-03-00~\cite{Lange:2001uf} package, while the \jpsi decay into the dielectron channel is performed using the EvtGen package coupled with the PHOTOS model~\cite{BARBERIO1994291} for the treatment of radiative decays ($\jpsi \rightarrow e^{+}e^{-}\gamma$). Prompt \jpsi are assumed to be unpolarised, while for the non-prompt \jpsi a small residual polarisation as predicted by EvtGen~\cite{Lange:2001uf} is considered. No further assumptions on the polarisation of both components are accounted, considering that existing measurements indicate small or no polarisation~\cite{ALICE:2018crw,ALICE:2020iev}.
The particle transport through the ALICE apparatus is handled by GEANT3~\cite{Brun:1082634}, considering a detailed description of the detector material and geometry. Detector responses and calibrations in MC simulations are tuned to data, taking into account time-dependent running conditions of all detectors included in the data acquisition.

The \pt spectrum of prompt \jpsi in simulations is tuned to match experimentally observed distributions, taking into account the centrality dependence. In particular, the fits to measured inclusive \jpsi yields from earlier publications~\cite{ALICE:2022inclJpsi,ALICE:2019nrq} are considered. To propagate the associated experimental uncertainties to the systematic uncertainties on acceptance-times-efficiency, all possible variations of the \pt shape within the envelope obtained by varying the fitting parameters according to their uncertainties are also considered.
For non-prompt \jpsi, different hypotheses are considered for the \pt distributions, including or excluding shadowing or suppression effects predicted by model calculations, such as those discussed in Sec.~\ref{Sec:discussion}. Taking into account all hypotheses discussed above for prompt and non-prompt \jpsi \pt spectra, an average correction factor for \fbp is evaluated.
As the acceptance-times-efficiency corrections are weakly dependent on \pt, the resulting correction applied on \fbp according to Eq.~\ref{eq:fBCorrProcedure} is small, being $\sim$5\% for the \pt integrated case, while for the different \pt intervals it ranges between 1\% to 3\%, depending on the width of the interval.

The resolution function, $F_{\rm prompt}(x)$ in Eq.~\ref{eq:signalPseudoPropDL}, defines the accuracy of $x$ in the reconstruction, and affects all different PDFs related to the pseudoproper decay length. It is described by the sum of two Gaussians and a symmetric power law function. It is determined as a function of both \pt and centrality from MC simulations. A tuning procedure is applied in MC simulations to minimise the residual discrepancy with respect to data in the average distance of closest approach (DCA) of the track to the reconstructed interaction vertex in a plane perpendicular to the beam direction. 
 A small centrality dependence is found for the $x$ resolution, in particular the RMS of the resolution function changes by about 5\% across different centrality classes, getting worse towards more central collisions. For the non-prompt \jpsi, the pseudoproper decay length distribution $F_{\rm B}(x)$ is modelled by the kinematic distribution of \jpsi from beauty-hadron decays obtained from the MC simulation described above, and convoluted with the resolution function.  
Finally, MC simulations are also employed to determine the shape of the invariant mass signal $M_{\rm Sig}(\mee)$, which is parametrised using a Crystal Ball function~\cite{Gaiser:1982yw}. Background PDFs, for both invariant mass and pseudoproper decay length distributions, are built from data. The invariant mass background shape $M_{\rm Bkg}(\mee)$, parametrised by a third order polynomial function, is estimated using the event-mixing technique, already used for the inclusive \jpsi analysis at midrapidity~\cite{ALICE:2022inclJpsi}, which also takes into account the residual background originating from correlated semileptonic decays of heavy-flavour hadrons.
\begin{figure}[t!]
  \includegraphics[width=0.99\textwidth]{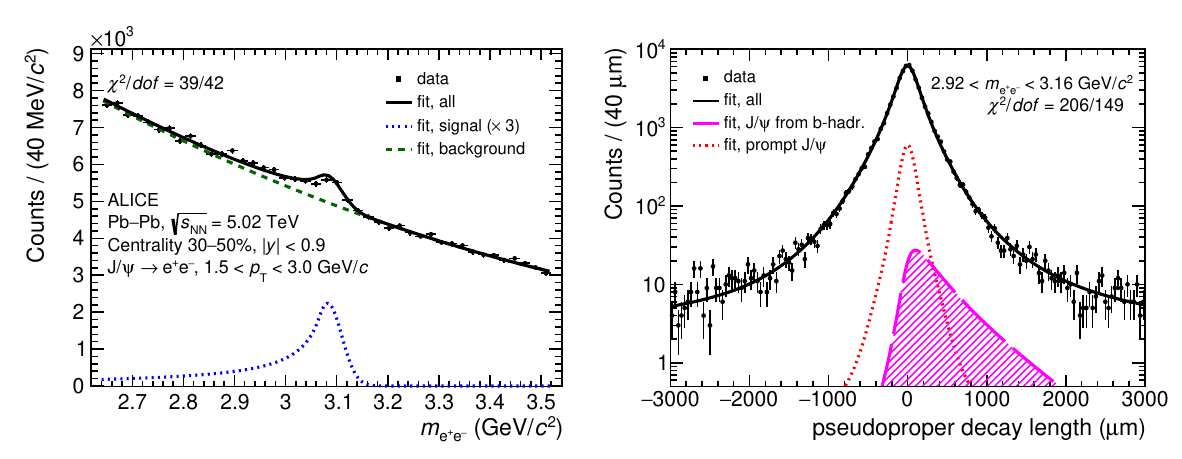}
        \caption{Invariant mass (left panel) and pseudoproper decay length (right panel) distributions of \jpsi candidates with maximum likelihood fit projections superimposed. The distributions in the figures correspond to selected candidates with $1.5 < \pt < 3.0$~\GeVc in the centrality class 30--50\%.  The pseudoproper decay length distribution is shown for \jpsi candidates reconstructed under the \jpsi mass peak, i.e. for $2.92 < m_{\rm ee} < 3.16$~GeV/$c^{2}$, for display purposes only. The $\chi^{2}$ values, obtained by comparing the binned distributions of data points and the corresponding projections of the total fit function, are also reported.
        }
  \label{Fig:LikelihoodFitProjection}
\end{figure}
\begin{table} [h!]
\centering
        \caption{ Invariant mass and \pt intervals considered for the definition of $F_{\rm Bkg}(x)$ in different centrality classes (see text for details). }
\resizebox{0.95\textwidth}{!}{
\begin{tabular}{ c|c|c }
 Centrality class & \mee\ sidebands (GeV/$c^{2}$) & \pt intervals (\GeVc) \\
 \hline
 0--10\% &  2.72$-$2.8, 2.8$-$2.9, 3.2$-$3.3, 3.3$-$3.4 & 1.5$-$3, 3$-$5, 5$-$7, 7$-$10 \\
 \hline
  10--30\%,30--50\%,50--90\% &  2.64$-$2.75, 2.75$-$2.85, 3.2$-$3.35, 3.35$-$3.52 & 1.5$-$3, 3$-$5, 5$-$10 \\
 \end{tabular}
        }
\label{tab:ptMassRanges}
\end{table}
For the pseudoproper decay length background function, $F_{\rm Bkg}(x)$, the same functional form and strategy described in Ref.~\cite{ALICE:2015nvt} is used. The function is determined for each centrality class by a fit to the data in four invariant mass regions on the sidebands of the \jpsi mass peak and in different \pt intervals, as summarised in Table~\ref{tab:ptMassRanges}. In each \pt interval, the $x$ background function in the invariant mass region which contains the nominal \jpsi mass value 
 is obtained by an interpolation procedure as the weighted combination of the PDFs determined in the other four invariant mass regions. The weights are chosen inversely proportional to the absolute or squared difference between the mean of the invariant mass distribution in the given mass interval and that in the interpolated region. Both hypotheses for the weights are considered for the study of the systematic uncertainty on the pseudoproper decay length background PDF. Additionally, the internal edges that define the invariant mass windows in Table~\ref{tab:ptMassRanges} are shifted by $\pm$~20~MeV/$c^2$, either in the same or opposite directions, and the interpolation procedure inside the signal region is repeated. For each \pt interval, a total of {\mbox 9 (variations of mass intervals)~$\times$ 2 (linear or quadratic weights) = 18} attempts are obtained for $F_{\rm Bkg}(x)$, which are used in turn in the unbinned likelihood fit procedure to determine \fbp. As the choice of a specific definition of mass intervals is in principle arbitrary, the central value of \fbp is computed considering the average of all the different attempts. Figure~\ref{Fig:LikelihoodFitProjection} shows an example of invariant mass (left panel) and pseudoproper decay length (right panel) distributions, corresponding to the centrality class 30--50\% and transverse momentum range $1.5 < \pt < 3.0$~\GeVc, with superimposed projections of the total maximum likelihood fit functions. The  non-prompt \jpsi fraction is measured in four (three) \pt intervals in the centrality class 0--10\% (10--30\% and 30--50\%), as well as in the \pt integrated case ($1.5 < \pt < 10$~\GeVc) in the centrality classes 0--10\%, 10--30\%, 30--50\%, and 50--90\%.

\subsubsection*{Systematic uncertainties}

\begin{table} [t!]
\centering
        \caption{Systematic uncertainties on \fb, expressed in \%, for the \pt integrated case, as well as for the lowest and the highest \pt interval in each centrality class.}

\resizebox{1.0\textwidth}{!}{
\begin{tabular}{ l|c|c|c|c|c|c|c|c|c|c }

          &  \multicolumn{3}{c|}{Cent. 0--10\% }      &   \multicolumn{3}{c|}{Cent. 10--30\% } &  \multicolumn{3}{c|}{Cent. 30--50\% } & Cent. 50--90\% \\
	\pt~(\GeVc)      & 1.5--10 & 1.5--3 & 7--10  & 1.5--10 & 1.5--3 & 5--10   & 1.5--10 & 1.5--3 & 5--10  & 1.5--10 \\
 \hline
        Resolution function &            2.5 & 5.0  & 1.0 &     3.5 & 5.0 & 1.0 &     2.5 & 3.5 & 1.0 &     2.5 \\
        \x\ PDF of background &          11.0 & 15.0 & 5.0 &   6.0 & 14.0 & 5.0 &    5.0 & 8.0 & 3.0 &     4.0 \\
        MC \pt\ distribution &                 6.0 & 1.5 &  3.0 &     6.0 & 1.5 & 4.0 &     6.0 & 1.5 &  3.0  &    6.0 \\
        \x\ PDF of non-prompt \jpsi &   2.0 & 2.0 &  1.5 &     2.0 & 4.0 &  1.0 &     2.0 & 2.5 & 1.0 &     3.0 \\
        $m_{\rm ee}$ PDF of signal &     0.5 & 1.5 & 0.5 &     0.5 & 0.5 & 0.5 &     0.5 & 0.5 & 0.5 &     0.5 \\
        $m_{\rm ee}$ PDF of background & 0.5 & 2.5 & 0.5 &     1.0 & 2.5 & 0.5 &     0.5 & 1.5 & 0.5 &     0.5  \\
 \hline
        Total &                          13.0 & 16.3 &  6.1 &   9.6 & 15.3 & 6.6 &     8.5 & 9.3 & 4.5 &    8.2 \\

 \end{tabular}
        }
\label{tab:fbsystable}
\end{table}

The systematic uncertainties on the measured non-prompt \jpsi fractions in different \pt intervals and centrality classes are summarised in Table~\ref{tab:fbsystable}. Most of the contributions are due to the incomplete knowledge of the different PDFs of \mee\ and $x$ in the likelihood fit. An additional contribution originates  from the assumptions of  the \pt distributions of \jpsi in MC simulations.

The dominant uncertainty is represented by the PDF of the pseudoproper decay length background, especially towards low \pt and more central collisions. Other contributions which become dominant in some cases are those originating from the resolution function, in particular in the lowest \pt intervals of each centrality, and the assumptions of the \pt distributions of \jpsi used in simulations, the latter particularly relevant for the \pt integrated case. 

The uncertainty on the resolution function is computed by considering the residual mismatch between data and MC simulations observed on the single track DCA resolution after the tuning procedure. The RMS of the resolution function is changed 
accordingly and the relative variation on \fb is taken as systematic uncertainty. The corresponding values are found to be larger at low \pt due to the worse pseudoproper decay length resolution and weakly dependent on centrality. 

The uncertainty due to the pseudoproper decay length background PDF is estimated by computing the RMS of non-prompt \jpsi fraction values obtained after changing the $F_{\rm Bkg}(x)$ in the likelihood fit, considering the eighteen attempts previously mentioned, which depend on the choice of the \mee\ sidebands. The uncertainty on \fb due to the $F_{\rm Bkg}(x)$ is the major contribution, especially at low-\pt and more central collisions where the signal-to-background ratio gets worse.

The systematic uncertainty related to the \pt distributions of prompt and non-prompt \jpsi mesons in MC simulations, mainly affecting the acceptance-times-efficiency correction on \fb according to Eq.~\ref{eq:fBCorrProcedure}, is evaluated by varying the prompt and non-prompt \jpsi \pt shape considering the variations previously discussed in this section. Taking into account all combinations, an average correction factor for \fbp is evaluated, while the maximum variation with respect to the average value is used to estimate the corresponding systematic uncertainty. This contribution increases for wider \pt intervals, reaching about 6\% for the \pt integrated interval, and is relatively independent of centrality. 

To quantify the systematic uncertainty related to the shape of the $x$ PDF of non-prompt \jpsi, the \pt distribution of non-prompt \jpsi is changed according to the same hypotheses considered for the estimate of the systematic uncertainty on the $\langle A \times \epsilon \rangle$ correction on \fb. Furthermore, the systematic uncertainty of the non-prompt \jpsi $x$ PDF originating from the description of the $h_{\rm b} \rightarrow \jpsi + X$ decay kinematic, with $h_{\rm b}$ representing any beauty mesons or baryons, is evaluated by considering PYTHIA 6.4 instead of EvtGen for decaying beauty hadrons. For both EvtGen and PYTHIA, all default decay modes of beauty hadrons available in the corresponding package and including a J/$\psi$ in the final state, are considered.

The uncertainty on the invariant mass PDF of the \jpsi signal is evaluated by changing the width of the Crystal Ball function in order to vary the fraction of candidates within the signal interval $2.92 < \mee < 3.16$~GeV/$c^{2}$ by $\pm$~2.5\%. The latter variation corresponds to the systematic uncertainty on the MC signal shape assigned in the inclusive \jpsi analysis~\cite{ALICE:2022inclJpsi}. 

The uncertainty related to the invariant mass background PDF is evaluated by taking the maximum variation obtained for \fb after using in the likelihood fit procedure alternative functions, namely fourth and fifth order polynomial functions, for fitting the event mixing based distribution used for the determination of $M_{\rm Bkg}(x)$.

The overall systematic uncertainty on \fb reaches a maximum of about 16\% in the lowest \pt interval and in 0--10\% centrality class, mostly as a consequence of both the increasing combinatorial background in more central collisions and the worsening of the $x$ resolution at low transverse momenta.

\subsection{Non-prompt \jpsi fractions in pp collisions at \s = 5.02 TeV}
\label{Sec:interpolation}

In pp collisions at \s = 5.02 TeV, non-prompt \jpsi fractions are measured at midrapidity by the ALICE collaboration~\cite{ALICE:2021edd} down to \pt~=~2~\GeVc. However, as the corresponding uncertainties are large and the \pt intervals in pp collisions do not match those of the \PbPb measurements, the \pt-differential \fb fractions are obtained via an interpolation procedure, already used in previous p--Pb and \PbPb analyses~\cite{ALICE:2015nvt,ALICE:2021lmn,ALICE:2022zig,ALICE:2018szk}. This procedure is based on available non-prompt \jpsi fraction measurements in pp and p$\overline{\rm p}$ collisions at midrapidity at several centre-of-mass energies, namely \s = 1.96 TeV (CDF~\cite{CDF:2004jtw}), 5.02 TeV (ALICE~\cite{ALICE:2021edd}, CMS~\cite{CMS:2017exb}), 7 TeV (ALICE~\cite{ALICE:2012vpz}, ATLAS~\cite{ATLAS:2015zdw,ATLAS:2011aqv}, CMS~\cite{CMS:2010nis}) and 8 TeV (ATLAS~\cite{ATLAS:2015zdw}).
In particular, using the semi-phenomenological function described in Ref.~\cite{ALICE:2015nvt}, which employs FONLL to describe the non-prompt \jpsi production cross section, the non-prompt \jpsi fraction in pp collisions at \s = 5.02 TeV ($f_{\rm B}^{\rm pp}$) as a function of \pt, needed for the computation of the reference for prompt and non-prompt \jpsi \RAA measurements, is derived via an energy interpolation. The average $f_{\rm B}^{\rm pp}$ in each \pt interval considered in the \PbPb analysis is obtained by reweighting the \pt-differential $f_{\rm B}^{\rm pp}$ values by the inclusive \pt-differential \jpsi cross section in pp collisions at \s = 5.02 TeV~\cite{ALICE:2019pid}.

\begin{table}[t!]
    \centering
        \caption{Fraction of non-prompt \jpsi with the corresponding uncertainties in pp collisions at \s = 5.02 TeV, obtained through the interpolation procedure, in different \pt intervals (see text for details).}
 \resizebox{0.40\textwidth}{!}{
	\begin{tabular}{ l|c }
        \pt (\GeVc) & $f_{\rm B}^{\rm pp}$ at \s = 5.02 TeV   \\
 \hline
  1.5--10 &  0.183 $\pm$ ~0.004  \\
  1.5--3 & 0.110 $\pm$ ~0.007    \\
  3--5 & 0.142 $\pm$ ~0.004      \\
  5--7 & 0.190 $\pm$ ~0.003      \\
  7--10 & 0.257 $\pm$ ~ 0.005    \\
  5--10 & 0.227 $\pm$ ~0.004     \\

\end{tabular}
}
\label{tab:tableFbPP}
\end{table}

The values of $f_{\rm B}^{\rm pp}$ in the \pt intervals considered in the \PbPb analysis are summarised in Table~\ref{tab:tableFbPP}. The uncertainty, which amounts to about 6.5\% in the lowest \pt interval and decreases to about 1.5\% at higher \pt, includes the contribution from experimental data and FONLL predictions, as well as the systematic uncertainty due to the choice of the functional form assumed for the energy interpolation (namely linear, exponential, and power law function).

\section{Results and discussion}
\label{Sec:discussion}

\subsection{ Non-prompt \jpsi fractions and \jpsi yields} 

\begin{figure}[t!]
  \includegraphics[width=0.52\textwidth]{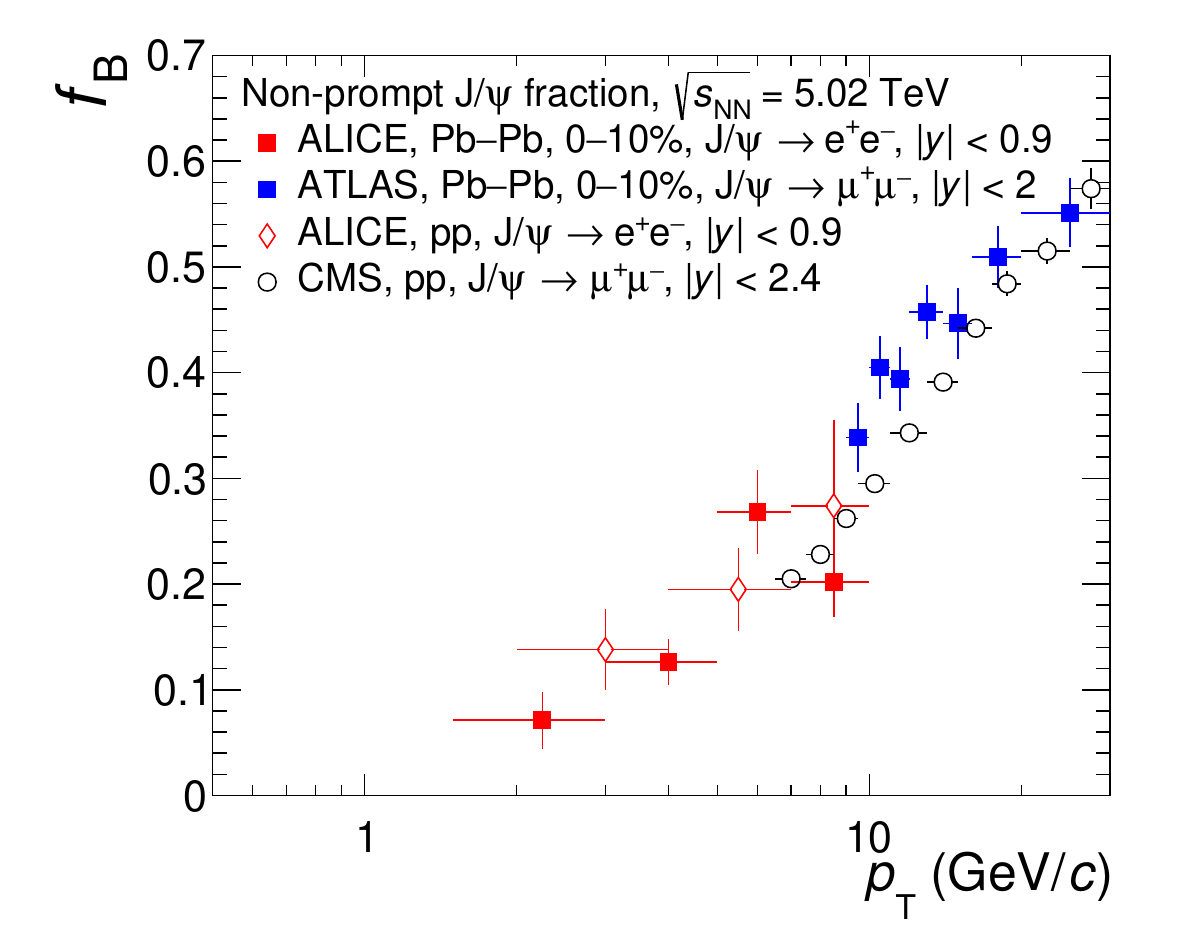}
  \includegraphics[width=0.52\textwidth]{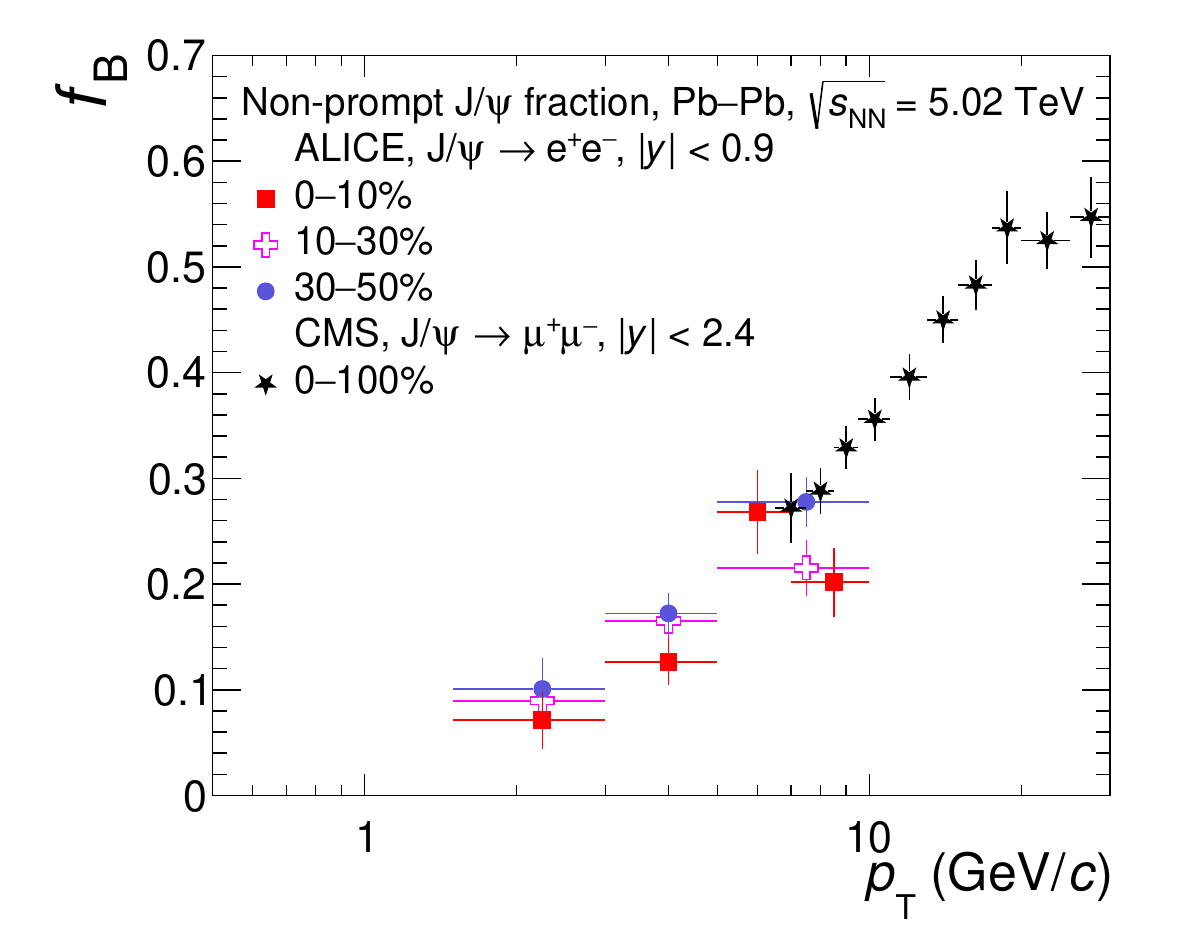}
	\caption{Non-prompt \jpsi fraction as a function of transverse momentum measured by the ALICE collaboration in 0--10\% most central Pb--Pb collisions (left panel) and in all centrality classes (right panel). Results in the left panel are compared with ATLAS midrapidity measurements~\cite{ATLAS:2018hqe} in the centrality class 0--10\%, and with ALICE~\cite{ALICE:2021edd}  and CMS~\cite{CMS:2017uuv} measurements in pp collisions at \s = 5.02 TeV. The ALICE measurements in the right panel are compared with similar midrapidity measurements from CMS~\cite{CMS:2017uuv}, performed in the centrality class 0--100\%. In both panels, the error bars represent the quadratic sum of statistical and systematic uncertainties.  }
  \label{Fig:fbVsPt}
\end{figure}

The non-prompt \jpsi fraction measured by the ALICE collaboration in 0--10\% Pb--Pb collisions as a function of \pt in $|y| < 0.9$ is shown in the 
left hand panel of Fig.~\ref{Fig:fbVsPt}. It is compared with similar results obtained by the ATLAS collaboration~\cite{ATLAS:2018hqe} at midrapidity, and available for \pt above 9.5~\GeVc.  
In the right hand panel, the \pt-differential non-prompt \jpsi fractions measured by the ALICE collaboration in different centrality classes are compared with CMS results~\cite{CMS:2017uuv} available for \pt $>$ 6.5 \GeVc in the centrality class 0--100\%. The ALICE results complement the existing high-\pt measurements from ATLAS and CMS, matching the decreasing trend observed from high towards low \pt. The results in the centrality interval 0--10\% suggest a smaller \fb\ compared to other centralities, in particular at low transverse momentum.
 In the left panel, non-prompt \jpsi fraction measurements in pp collisions at \s = 5.02 TeV obtained by the ALICE~\cite{ALICE:2021edd} and CMS~\cite{CMS:2017uuv} collaborations are also shown for comparison, and exhibit a trend similar to the one observed in \PbPb collisions. In the \pt range 10 to 20 \GeVc, the non-prompt \jpsi fractions are clearly higher in \PbPb compared to pp collisions, possibly suggesting a stronger nuclear suppression of prompt charmonia compared to beauty hadrons. On the other hand, at low-\pt results in the two systems exhibit similar values within uncertainties. 

\begin{figure}[t!]
\centering
  \includegraphics[width=0.59\textwidth]{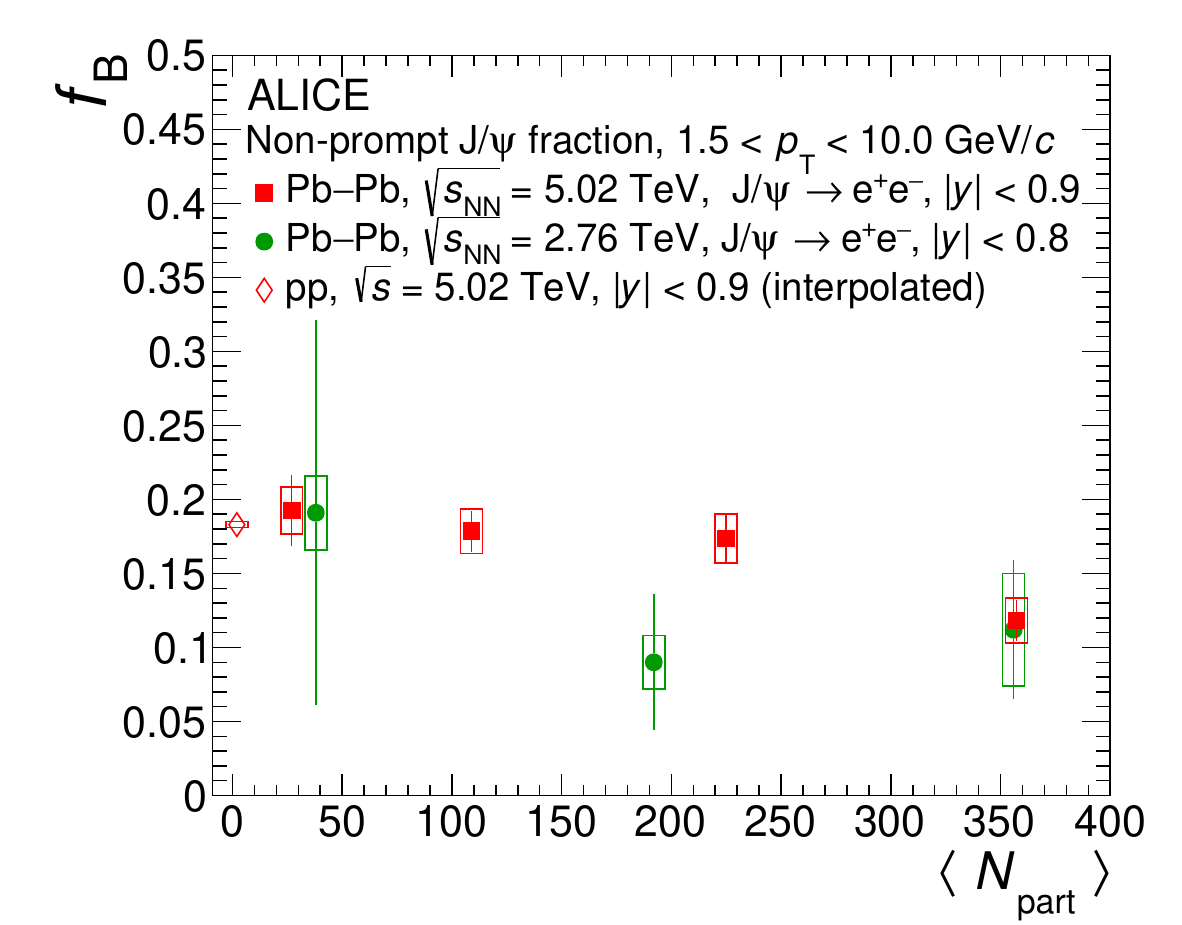}
	\caption{Centrality dependence (expressed in terms of average number of participants) of the non-prompt \jpsi fraction measured by ALICE in \PbPb collisions at \snn = 5.02 TeV in the transverse momentum interval 1.5 $<$ \pt $<$ 10 \GeVc. Results are compared with previous ALICE measurements performed in \PbPb collisions at \snn = 2.76 TeV~\cite{ALICE:2015nvt}. The \pt-integrated non-prompt \jpsi fraction in pp collisions at \s = 5.02 TeV obtained from an interpolation procedure (see text for details) is also shown. Statistical and systematic uncertainties are shown by error bars and boxes, respectively. 
        }
  \label{Fig:fbVsCentrality}
\end{figure}

Figure~\ref{Fig:fbVsCentrality} presents the non-prompt \jpsi fraction as a function of centrality, expressed in terms of the average number of participants $\langle N_{\rm part} \rangle$, measured by ALICE in \PbPb collisions at \snn = 5.02 TeV in the transverse momentum interval 1.5 $<$ \pt $<$ 10 \GeVc. Within uncertainties, no centrality dependence is observed between peripheral and semicentral collisions, while the measured \fb\ in 0--10\% most central collisions decreases in comparison to an average value computed by considering all other centralities with a significance of about 2.5$\sigma$ considering both statistical and systematic uncertainties. This is consistent with the observations for the \pt-dependent results shown in Fig.~\ref{Fig:fbVsPt} and is compatible with the hypothesis of a strong contribution of prompt \jpsi originating from regeneration, which is expected to be larger in central compared to peripheral collisions. Current results are compared with earlier ALICE measurements based on \PbPb collisions at \snn = 2.76 TeV~\cite{ALICE:2015nvt}. 
The statistical precision is significantly improved thanks to the larger event sample available from LHC Run 2. 
In the same figure, the \pt-integrated non-prompt \jpsi fraction in pp collisions at \s = 5.02 TeV obtained by the interpolation procedure described in Section~\ref{Sec:interpolation} is also shown. The corresponding value is found to be comparable with the non-prompt \jpsi fractions measured in peripheral and semicentral Pb--Pb collisions.     

The \pt-differential production yields of prompt and non-prompt \jpsi in a given centrality interval are computed by combining the non-prompt \jpsi fractions with the measured yields of inclusive \jpsi~\cite{ALICE:2022inclJpsi}, ${\ensuremath{\mathrm{d}N^{\rm incl.~\jpsi}/\mathrm{d}y\mathrm{d}\pt}\xspace}$, as

\begin{equation}
	{\ensuremath{\frac{\mathrm{d}N^{\pJPsi \leftarrow {\it h}_{\rm B}}}{\mathrm{d}y\mathrm{d}\pt}}\xspace}  = \fb \times  {\ensuremath{\frac{\mathrm{d}N^{\rm incl.~\jpsi}}{\mathrm{d}y\mathrm{d}\pt}}\xspace},~~~~~~~~~~~~~~~~~~{\ensuremath{\frac{\mathrm{d}N^{\rm prompt~\jpsi}}{\mathrm{d}y\mathrm{d}\pt}}\xspace}  = (1-\fb) \times  {\ensuremath{\frac{\mathrm{d}N^{\rm incl.~\jpsi}}{\mathrm{d}y\mathrm{d}\pt}}\xspace}. 
        \label{eq:yields010}
\end{equation}

\begin{figure}[t!]
\centering
  \includegraphics[width=0.70\textwidth]{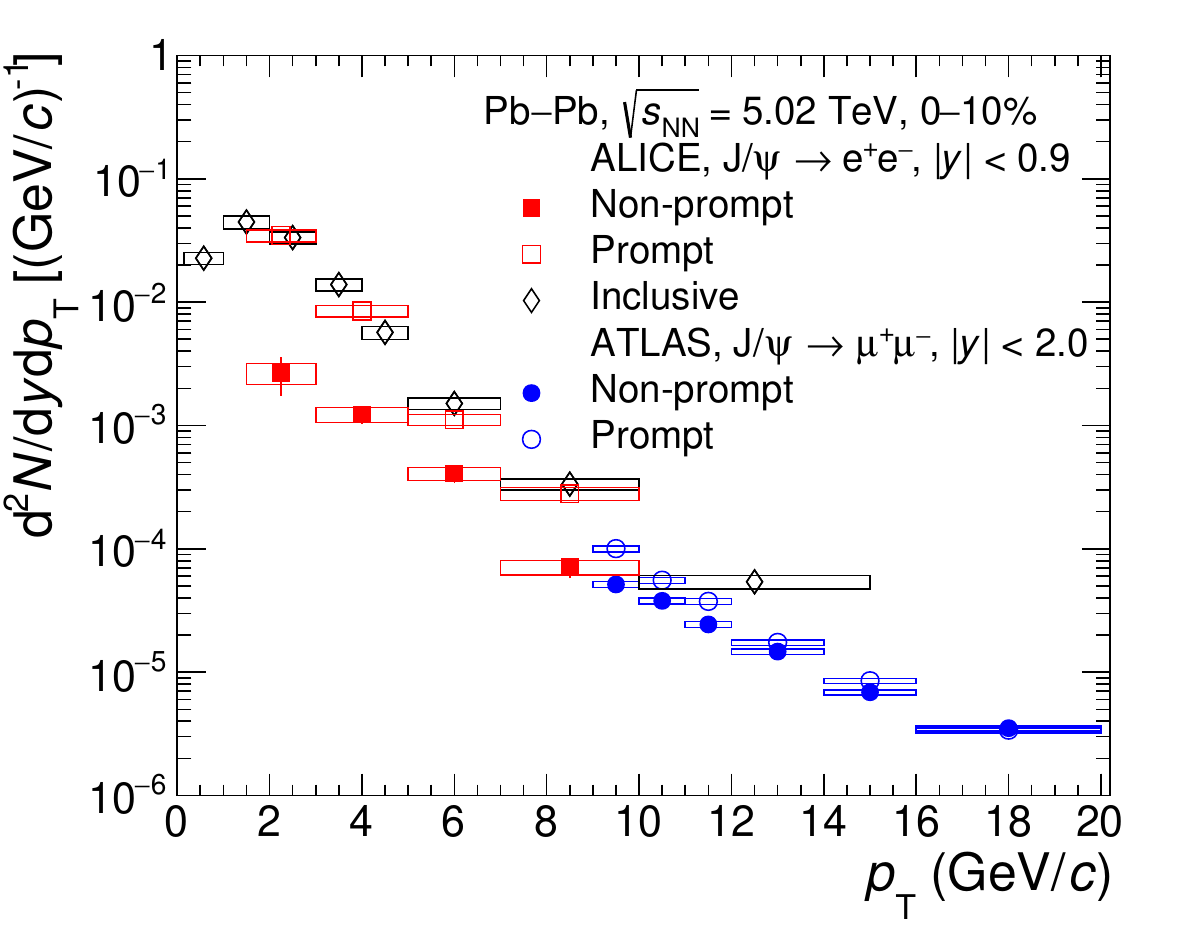}
	\caption{Prompt and non-prompt \jpsi \pt-differential production yields measured by ALICE in the 0--10\% centrality class at midrapidity, compared with similar measurements from the ATLAS collaboration~\cite{ATLAS:2018hqe} in the same centrality class. Inclusive \jpsi yields measured by the ALICE collaboration in 0--10\%~\cite{ALICE:2022inclJpsi} are shown for comparison. 
	Error bars and boxes represent statistical and systematic uncertainties, respectively. 
        }
  \label{Fig:yields010}
\end{figure}

Figure~\ref{Fig:yields010} shows the \pt-differential production yields of prompt and non-prompt \jpsi in the centrality interval 0--10\%. Statistical and systematic uncertainties on prompt and non-prompt \jpsi yields are evaluated by adding in quadrature the corresponding uncertainties on \fb and inclusive \jpsi yields, the latter discussed in detail in Ref.~\cite{ALICE:2022inclJpsi}. The statistical uncertainties of inclusive \jpsi yields vary from about 5\% to 10\%, depending on the \pt and the centrality of the collision.
Regarding the systematic uncertainty of the inclusive \jpsi yield, it ranges within 10--12\% (8--10\%) in the centrality class 0--10\% (30--50\%) and shows no significant \pt dependence. 
The measurements from ALICE are compared with \pt-differential production yields obtained by the ATLAS collaboration~\cite{ATLAS:2018hqe}, as well as with inclusive \jpsi measurements performed by the ALICE collaboration~\cite{ALICE:2022inclJpsi}. As already observed for the non-prompt \jpsi fraction, the ALICE measurements provide complementary \pt coverage to ATLAS results, and they extend the measurement down to \pt = 1.5 \GeVc. In addition, the ALICE and ATLAS results are qualitatively compatible in the overlapping region and show similar slopes, which result in an overall smooth trend over the full \pt range. Given the relatively small non-prompt \jpsi fraction below 10 \GeVc, the inclusive \jpsi yield and that of the prompt \jpsi, both measured by the ALICE collaboration, are found to be comparable. However, the inclusive \jpsi yield is measured in finer \pt intervals and subtends a wider range in \pt. 

\subsection{\jpsi nuclear modification factors}

The nuclear modification factor \RAA of prompt and non-prompt \jpsi is obtained by combining the \RAA of inclusive \jpsi~\cite{ALICE:2022inclJpsi}, with the non-prompt \jpsi fractions measured in \PbPb collisions at \snn = 5.02 TeV normalised to those in pp collisions obtained at the same centre-of-mass energy through the interpolation procedure described in Section~\ref{Sec:interpolation}:

\begin{equation}
        \RAA^{\jpsi \leftarrow {\it h}_{\rm B}} = \frac{f_{\rm B}^{\rm Pb-Pb}}{f_{\rm B}^{\rm pp}} \times \RAA^{\rm inclusive~\pJPsi},~~~~~~~~~~~~~~~~~~\RAA^{\rm prompt~\jpsi} = \frac{1-f_{\rm B}^{\rm Pb-Pb}}{1-f_{\rm B}^{\rm pp}} \times \RAA^{\rm inclusive~\pJPsi}. 
        \label{eq:raaPromptAndNonPrompt}
\end{equation}

Statistical and systematic uncertainties on prompt and non-prompt \jpsi \RAA are obtained by adding in quadrature the corresponding uncertainties on inclusive \jpsi \RAA~\cite{ALICE:2022inclJpsi} and non-prompt \jpsi fractions in Pb--Pb and in pp collisions, assuming all of them uncorrelated from each other. Uncertainties on inclusive \jpsi \RAA, both statistical and systematic, include the corresponding contributions due to inclusive \jpsi yields previously mentioned and inclusive \jpsi cross section in pp collisions at \s = 5.02 TeV~\cite{ALICE:2019pid}, the latter used as reference in the inclusive \jpsi \RAA computation. The statistical uncertainty on the pp reference cross section increases with \pt from about 7\% to 18\%, while the systematic uncertainty is poorly dependent on \pt and it amounts to about 6\%. The latter value do not include the normalisation uncertainty due to luminosity determination, which is 2.2\% and it is considered within the global uncertainties when computing \RAA. 

\begin{figure}[t!]
  \includegraphics[width=0.52\textwidth]{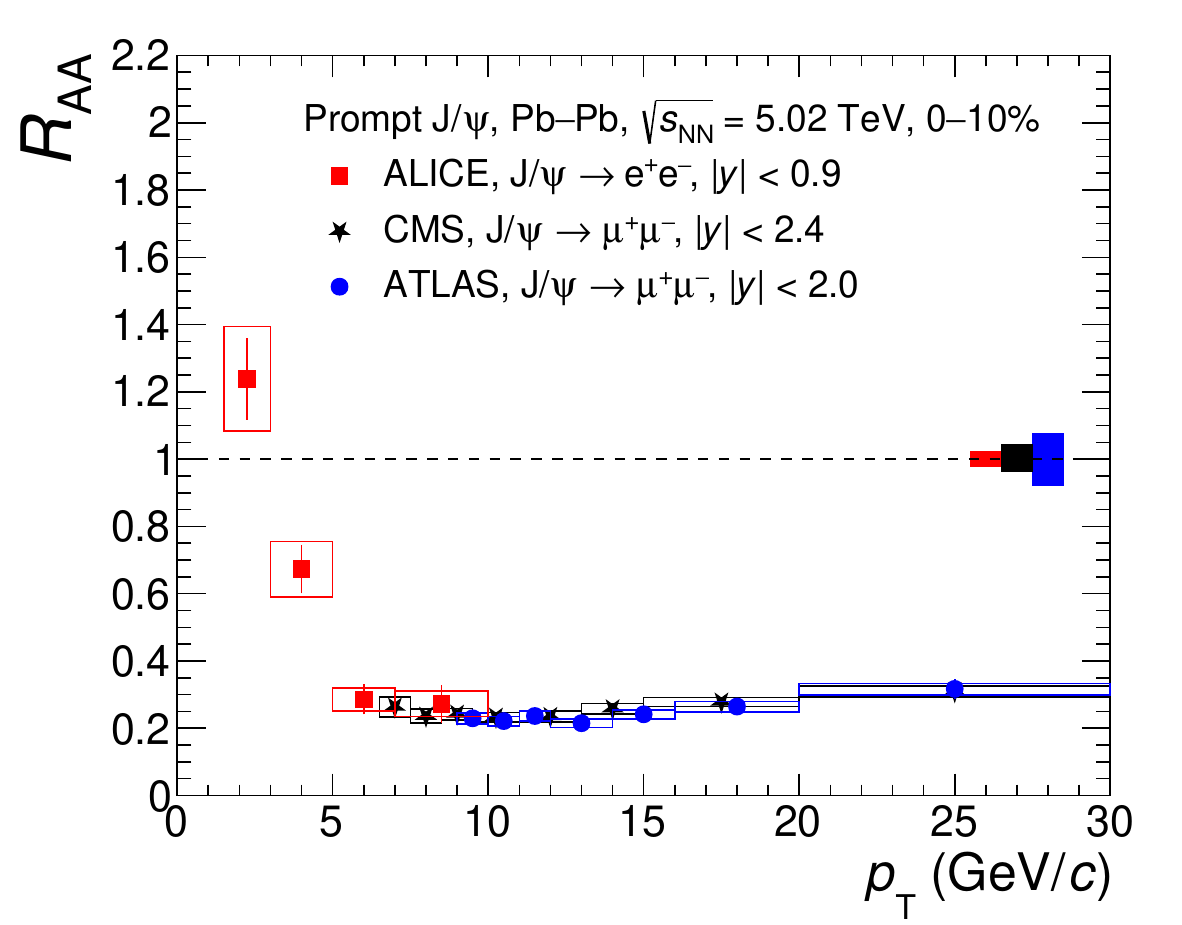}
  \includegraphics[width=0.52\textwidth]{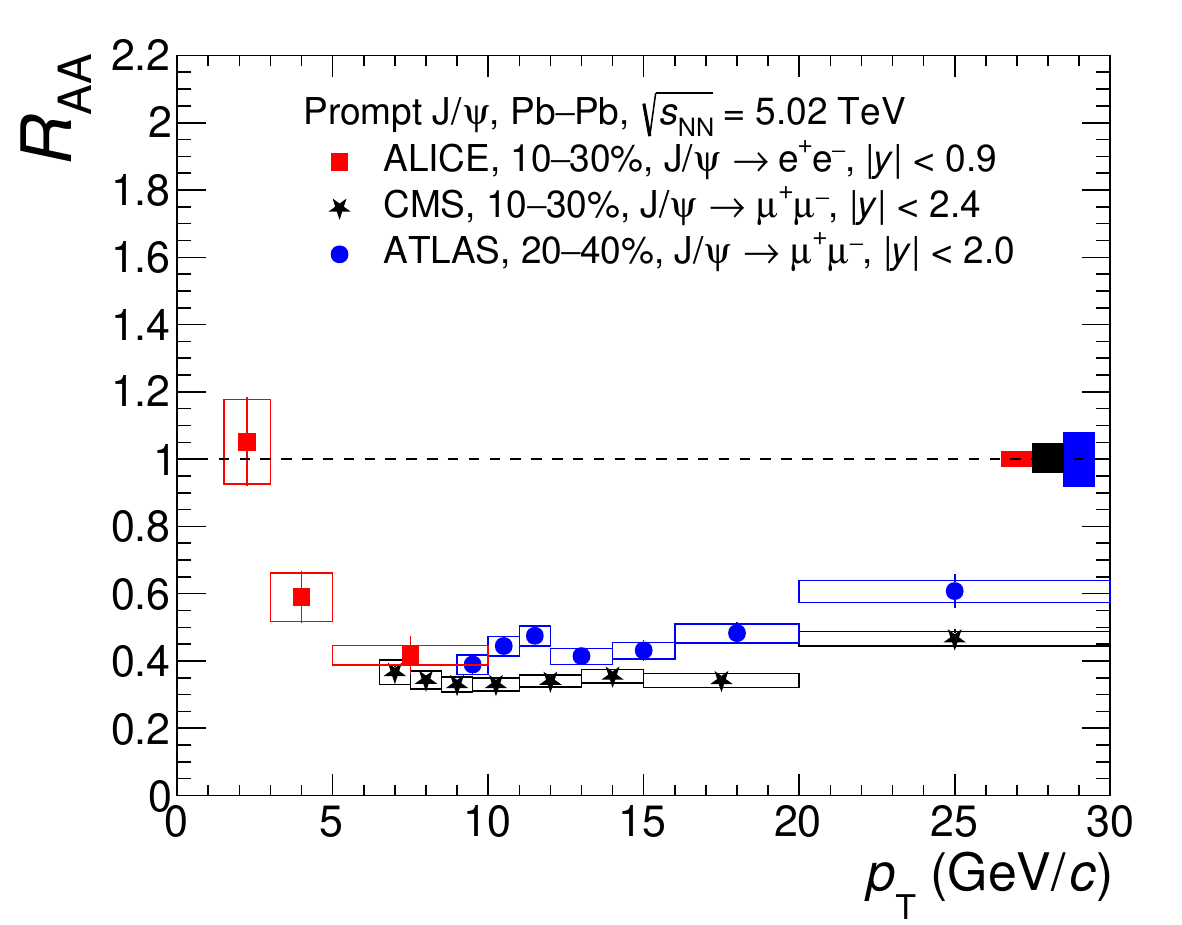} \\
  \includegraphics[width=0.52\textwidth]{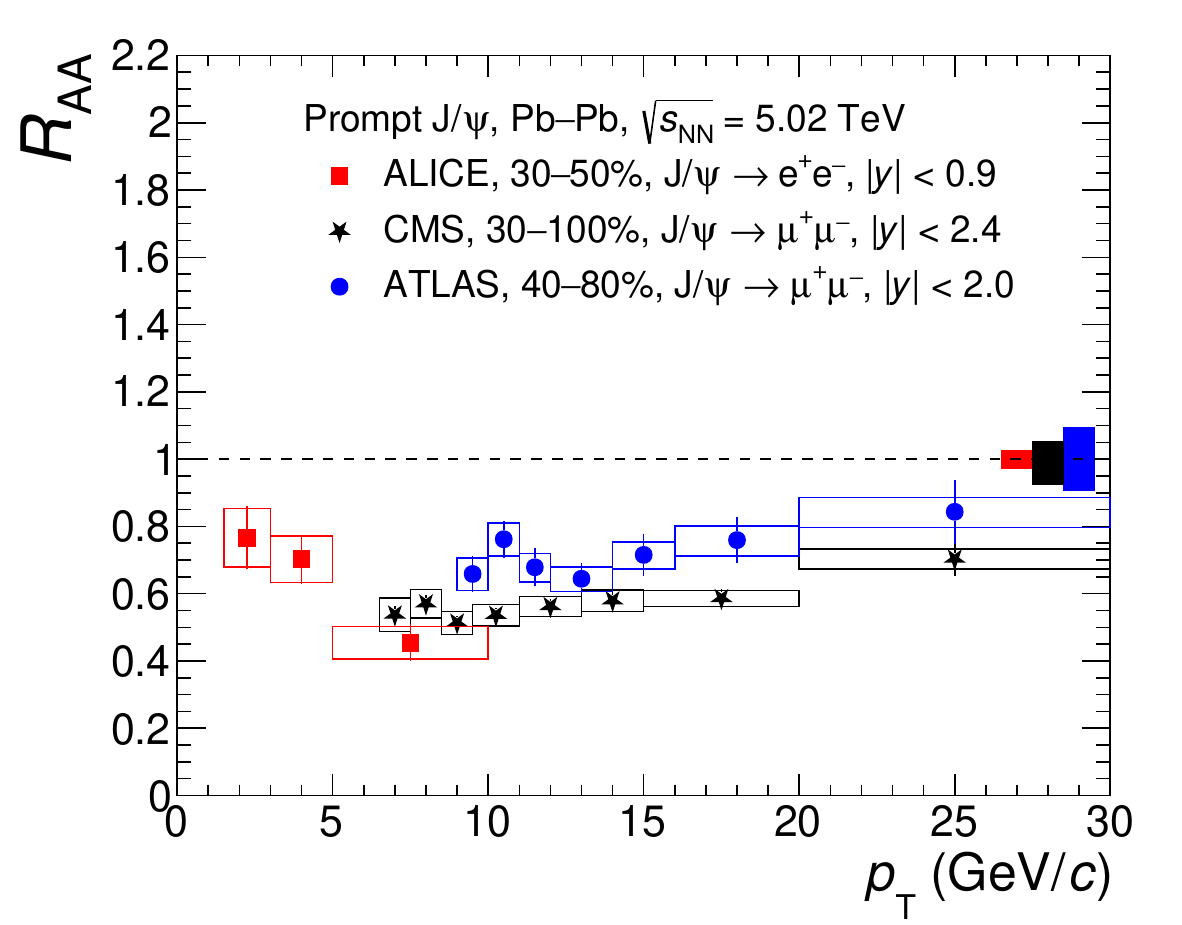}
  \includegraphics[width=0.52\textwidth]{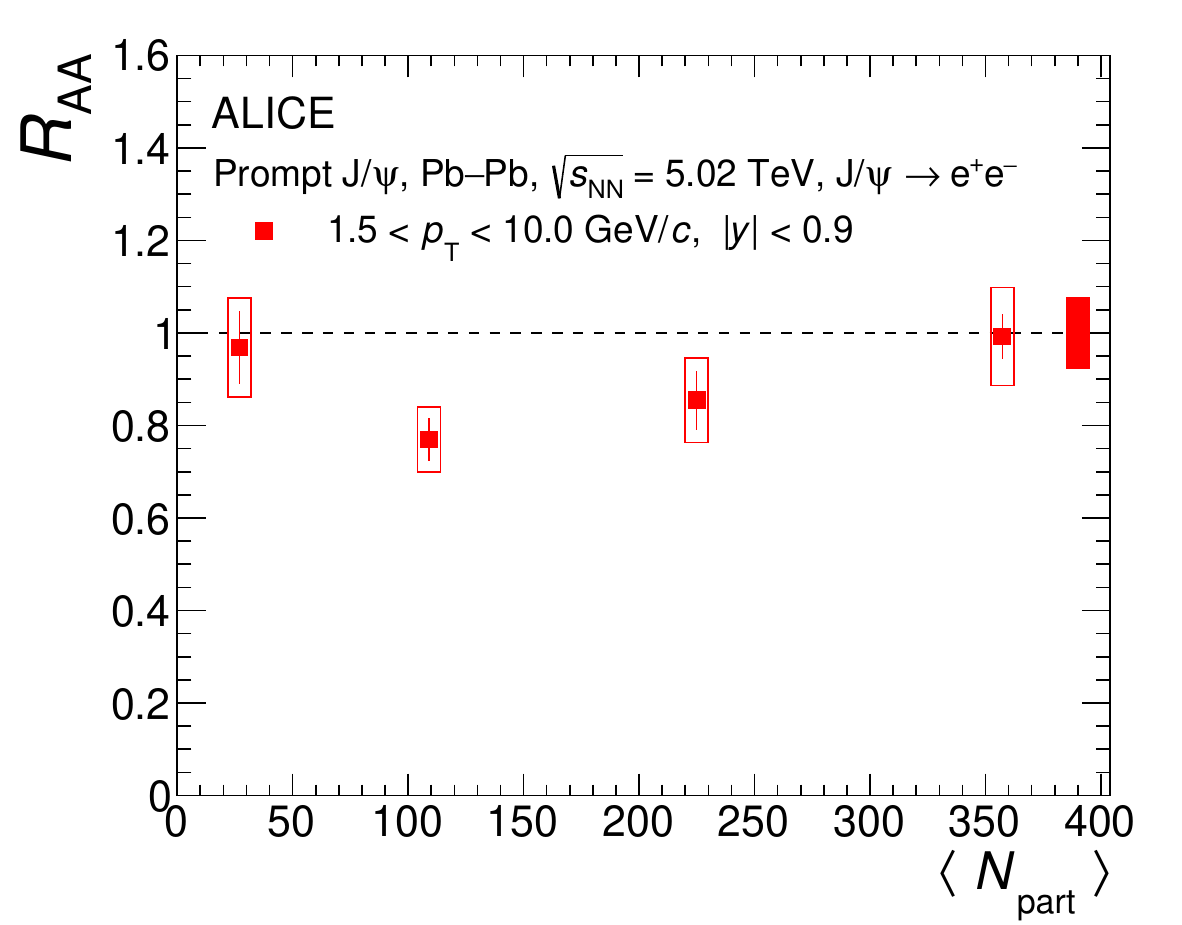} 
        \caption{Nuclear modification factor of prompt \jpsi as a function of \pt in 0--10\% (upper left panel), 10--30\% (top right panel) and 30--50\% (bottom left panel) centrality classes. Results are compared with similar measurements from the ATLAS~\cite{ATLAS:2018hqe} and CMS~\cite{CMS:2017uuv} collaborations. The centrality dependent prompt \jpsi \RAA in 1.5 $<$ \pt $<$ 10 \GeVc is shown in the bottom right panel (centrality is expressed in terms of average number of participants).  Error bars and boxes represent, respectively, statistical and systematic uncertainties uncorrelated with \pt (centrality, for the bottom right panel). Global uncertainties are shown as boxes around unity.
        }
  \label{Fig:PromptRaaVsPTComparisonWithExp}
\end{figure}

\begin{figure}[h!]
  \includegraphics[width=0.52\textwidth]{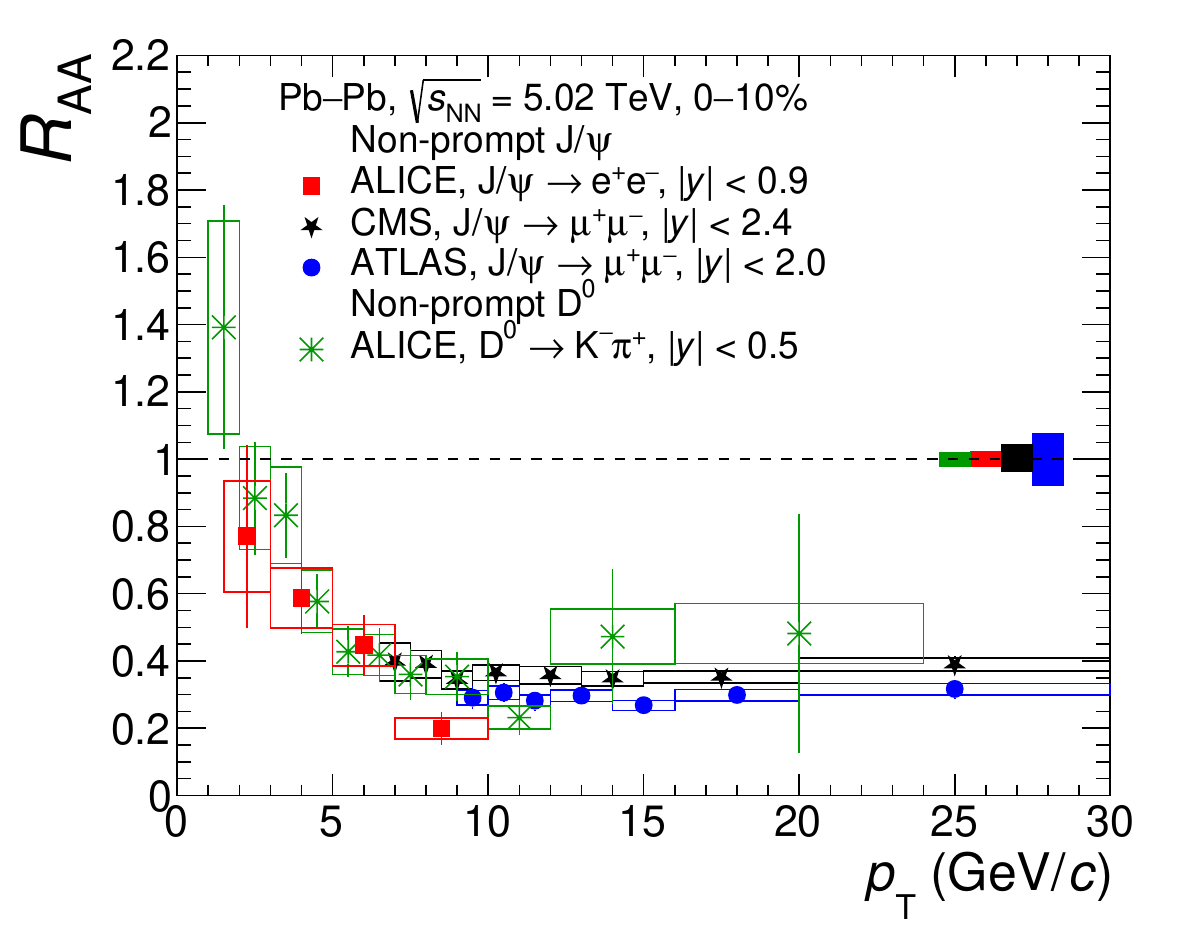}
  \includegraphics[width=0.52\textwidth]{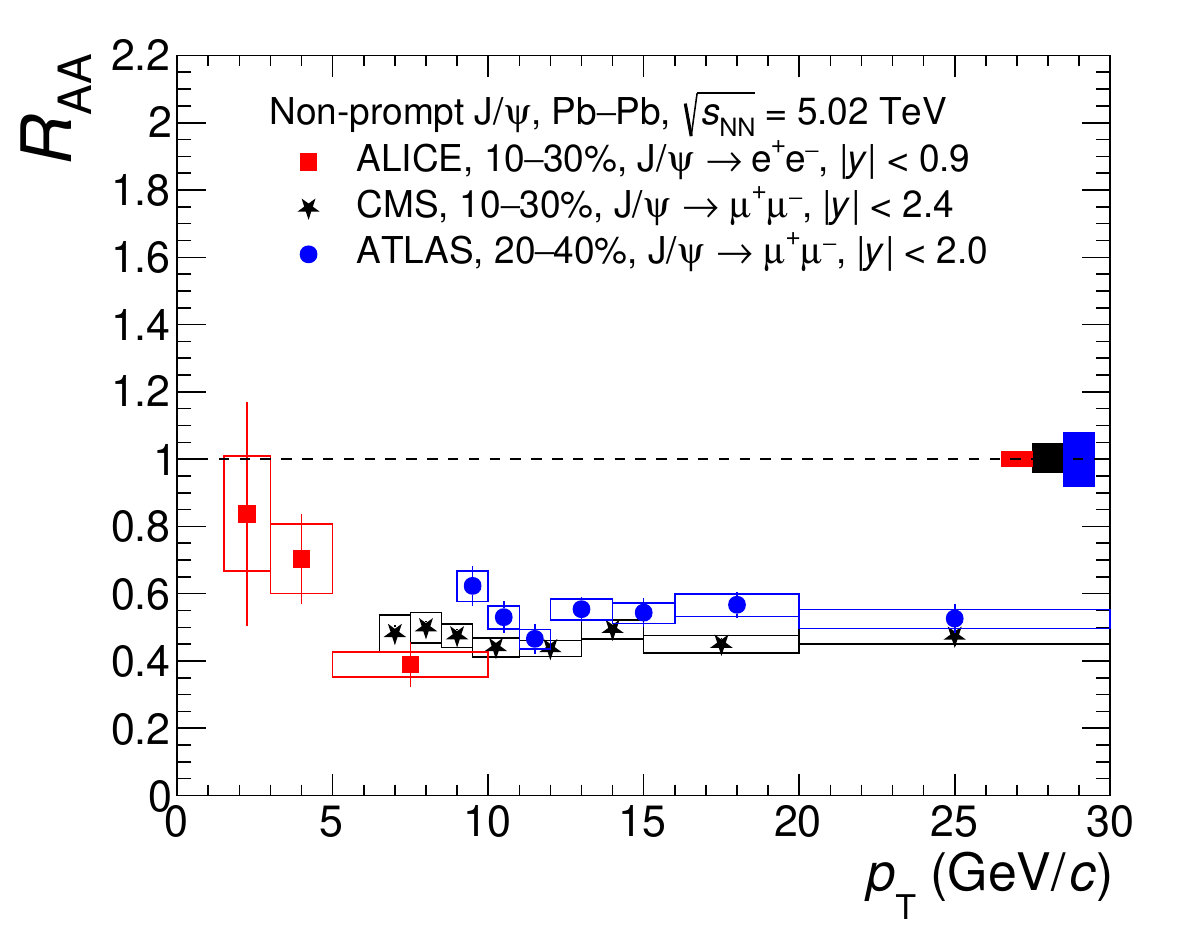} \\
  \includegraphics[width=0.52\textwidth]{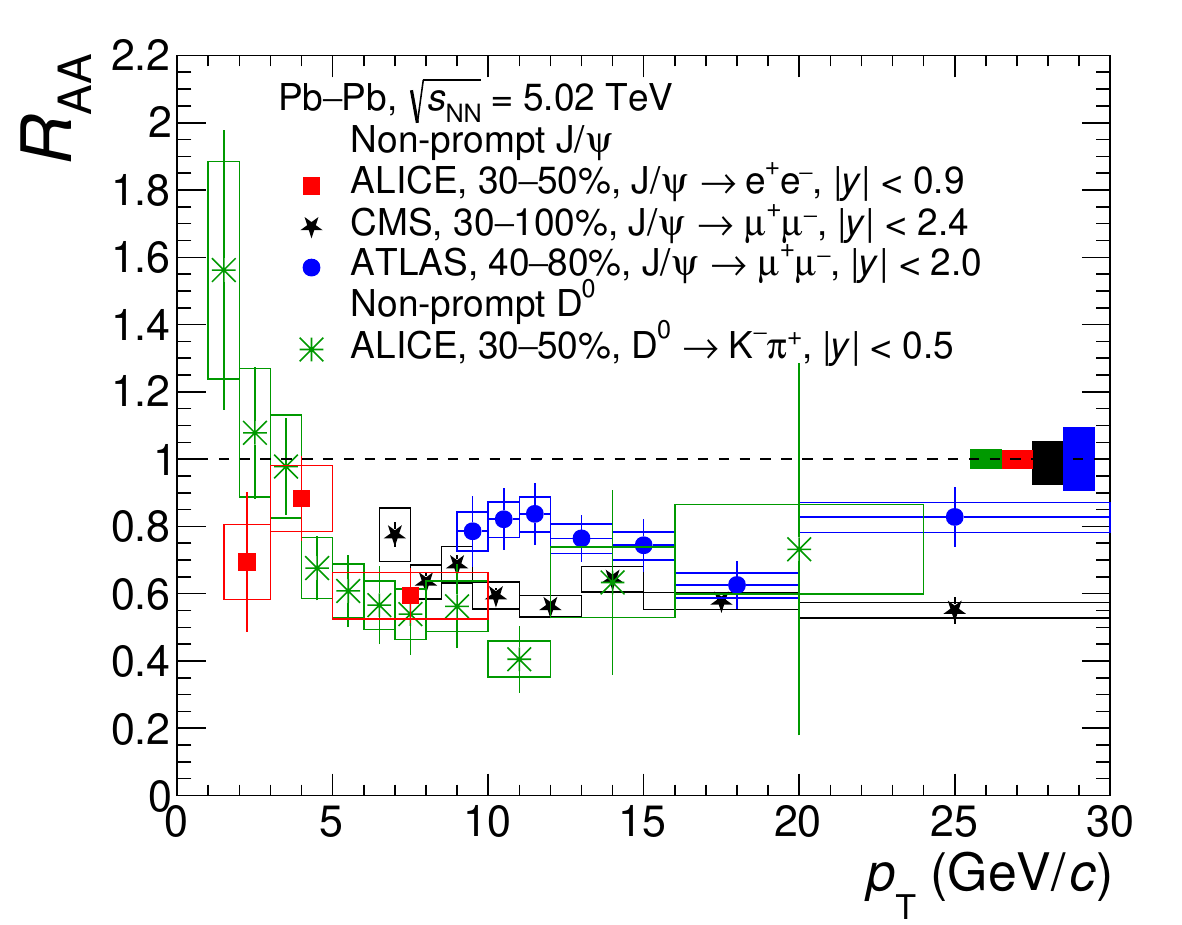}
  \includegraphics[width=0.52\textwidth]{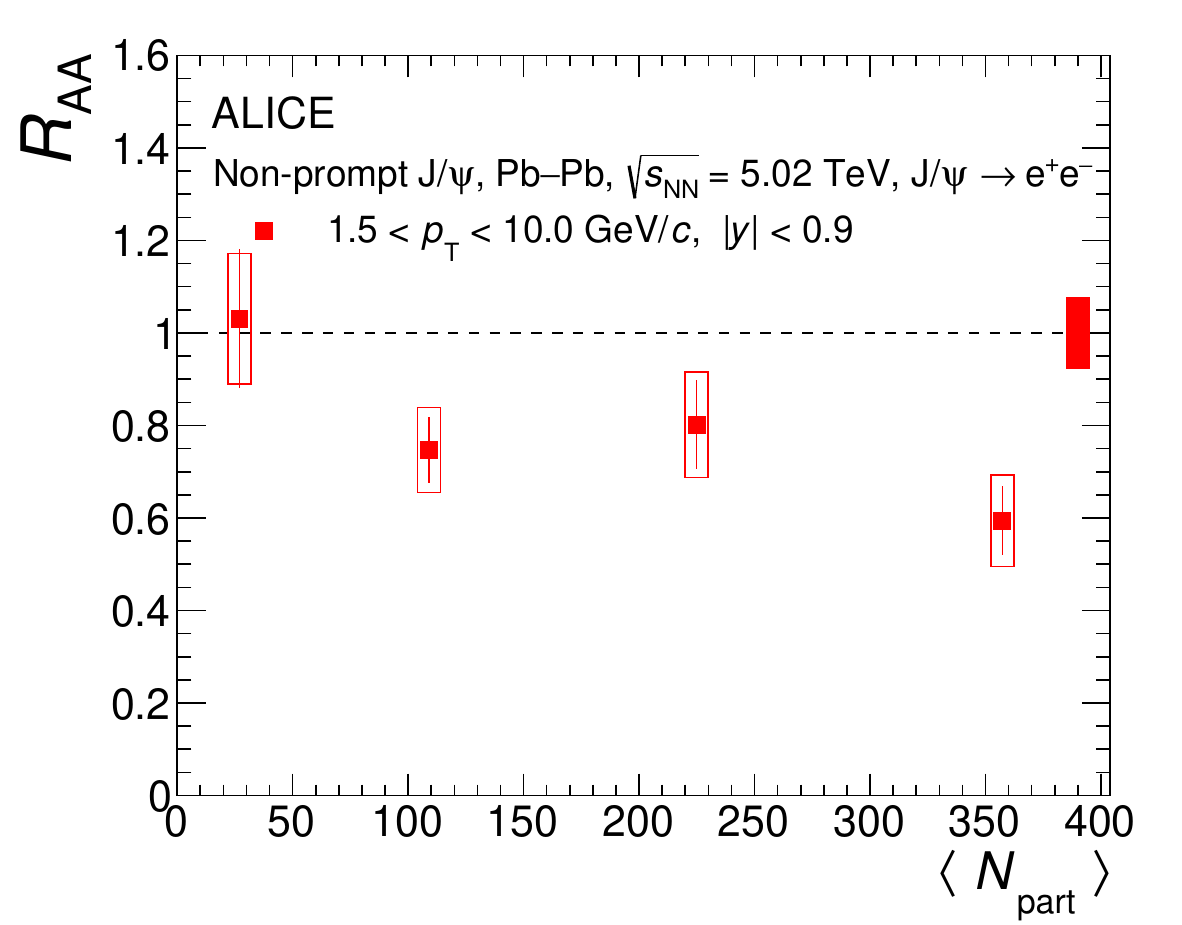} 
        \caption{Nuclear modification factor of non-prompt \jpsi as a function of \pt in the 0--10\% (upper left panel), 10--30\% (top right panel) and 30--50\% (bottom left panel). Results are compared with similar measurements from the ATLAS~\cite{ATLAS:2018hqe} and CMS~\cite{CMS:2017uuv} collaborations. Results in 0--10\% and 30--50\% are also compared to non-prompt D$^{0}$ \RAA measurements~\cite{ALICE:2022tji} in the same centrality classes. The centrality dependent non-prompt \jpsi \RAA in 1.5 $<$ \pt $<$ 10 \GeVc is shown in the bottom right panel (centrality is expressed in terms of average number of participants). Error bars and boxes represent statistical and systematic uncertainties uncorrelated with \pt (centrality, for the bottom right panel). Global uncertainties are shown as boxes around unity.
        }
  \label{Fig:NonPromptRaaVsPTComparisonWithExp}
\end{figure}

The \pt-differential nuclear modification factor of prompt \jpsi is shown in Fig.~\ref{Fig:PromptRaaVsPTComparisonWithExp} in different centrality classes, namely 0--10\% (top left panel), 10--30\% (top right panel) and 30--50\% (bottom left panel). Global uncertainties, shown as boxes around unity, include correlated uncertainties on the pp reference due to normalisation as well as on the $\langle \TAA \rangle$~\cite{ALICE:2022inclJpsi}. The latter uncertainty decreases with the collision centrality, varying from 0.7\% in 0--10\% most central collisions to 2\% in the 50--90\% centrality class. The \RAA of prompt \jpsi in 1.5 $<$ \pt $<$ 10 \GeVc and as a function of centrality is shown in the bottom right panel of the same figure. The global uncertainty includes the contributions from the inclusive \jpsi cross section and from $f_{\rm B}^{\rm pp}$ in pp collisions, both integrated over \pt (1.5 $<$ \pt $<$ 10 \GeVc). For \pt $>$ 5 \GeVc, the prompt \jpsi \RAA in the centrality classes 10--30\% and 30--50\% reaches a value of about 0.4, while in 0--10\% collisions it decreases to about 0.2, indicating a stronger suppression. The prompt \jpsi \RAA increases towards low \pt and it exceeds unity in the lowest \pt interval (1.5$-$3~\GeVc) of the 0--10\% most central collisions. As a consequence, the \pt integrated prompt \jpsi \RAA (1.5 $<$ \pt $<$ 10 \GeVc) also rises towards most central collisions, as shown in the bottom right panel of Fig.~\ref{Fig:PromptRaaVsPTComparisonWithExp}. According to prompt \jpsi measurements performed by the ALICE collaboration in p--Pb collisions at \snn = 5.02 TeV~\cite{ALICE:2021lmn}, the \RpPb is found to be lower than unity within $1 < \pt < 3$~\GeVc, suggesting significant CNM effects at play in Pb--Pb collisions in this transverse momentum region, while it becomes compatible with unity for $\pt > 3$~\GeVc.
The results from the ALICE collaboration are compared with similar measurements carried out by the ATLAS~\cite{ATLAS:2018hqe} and CMS~\cite{CMS:2017uuv} collaborations in the rapidity intervals $|y| < 2.0$ and $|y| < 2.4$, respectively. In the centrality interval 0--10\%, the results from CMS and ATLAS are in good agreement with the ALICE measurements in the overlapping \pt interval, while in semicentral collisions the agreement is better with CMS than with ATLAS. It is worth noting that a smaller suppression for ATLAS measurements is expected in semicentral events, as the corresponding results use a significantly more peripheral collision sample compared to ALICE and CMS. 

Figure~\ref{Fig:NonPromptRaaVsPTComparisonWithExp} shows the nuclear modification factor of non-prompt \jpsi as a function of \pt in the same centrality classes discussed previously for the prompt \jpsi \RAA, namely 0--10\%, 10--30\% and 30--50\%.  The boxes around unity represent global uncertainties due to the normalisation of the pp reference cross section and $\langle \TAA \rangle$.  Unlike the prompt \jpsi \RAA, the \pt-differential nuclear modification factors of non-prompt \jpsi are similar in the different centrality classes within experimental uncertainties below 5 \GeVc, while at higher \pt the suppression is larger in 0--10\% most central collisions. The \pt-integrated \RAA in the bottom right panel of Fig.~\ref{Fig:NonPromptRaaVsPTComparisonWithExp} hints at a decreasing trend towards more central collisions, reaching an \RAA value of about 0.6\ in the 0--10\% centrality class. In the case of non-prompt \jpsi, CNM effects are found to be small in the full measured \pt interval, as non-prompt \jpsi \RpPb measurements at \snn = 5.02 TeV~\cite{ALICE:2021lmn} are everywhere compatible with unity, although uncertainties are large. Therefore, the observed modification can be attributed to the QGP formation in \PbPb collisions. Results as a function of \pt are shown together with non-prompt \jpsi \RAA measurements from the ATLAS~\cite{ATLAS:2018hqe} and CMS~\cite{CMS:2017uuv} collaborations, which are in good agreement with ALICE measurements in the overlapping \pt region. In 0--10\% most central collisions, the \RAA from ALICE in the highest \pt interval shows a small tension of about 1.4$\sigma$ and 2.5$\sigma$ with respect to ATLAS and CMS, respectively. 
The non-prompt \jpsi \RAA measurements in 0--10\% and 30--50\% centrality classes are also compared with non-prompt D$^{0}$ results~\cite{ALICE:2022tji} measured at midrapidity by the ALICE collaboration. The results are compatible within uncertainties, despite possible differences that might originate from different decay kinematics of beauty hadrons to \jpsi and D$^{0}$.

\subsection{Comparison with models for prompt \jpsi production}

In the following, prompt \jpsi measurements are compared with different phenomenological models in relativistic heavy-ion collisions. 

\begin{figure}[t!]
  \includegraphics[width=0.51\textwidth]{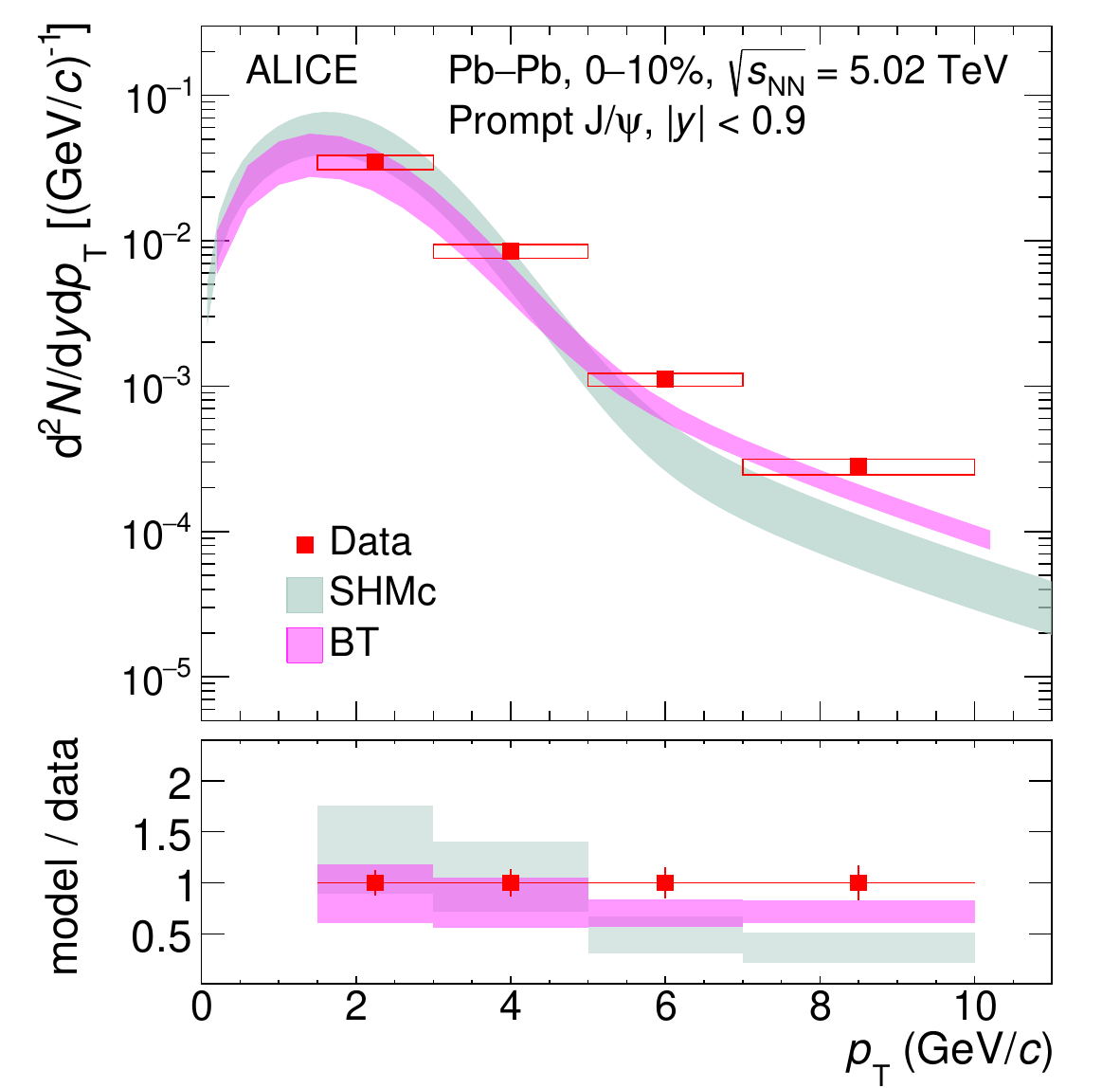}
  \includegraphics[width=0.51\textwidth]{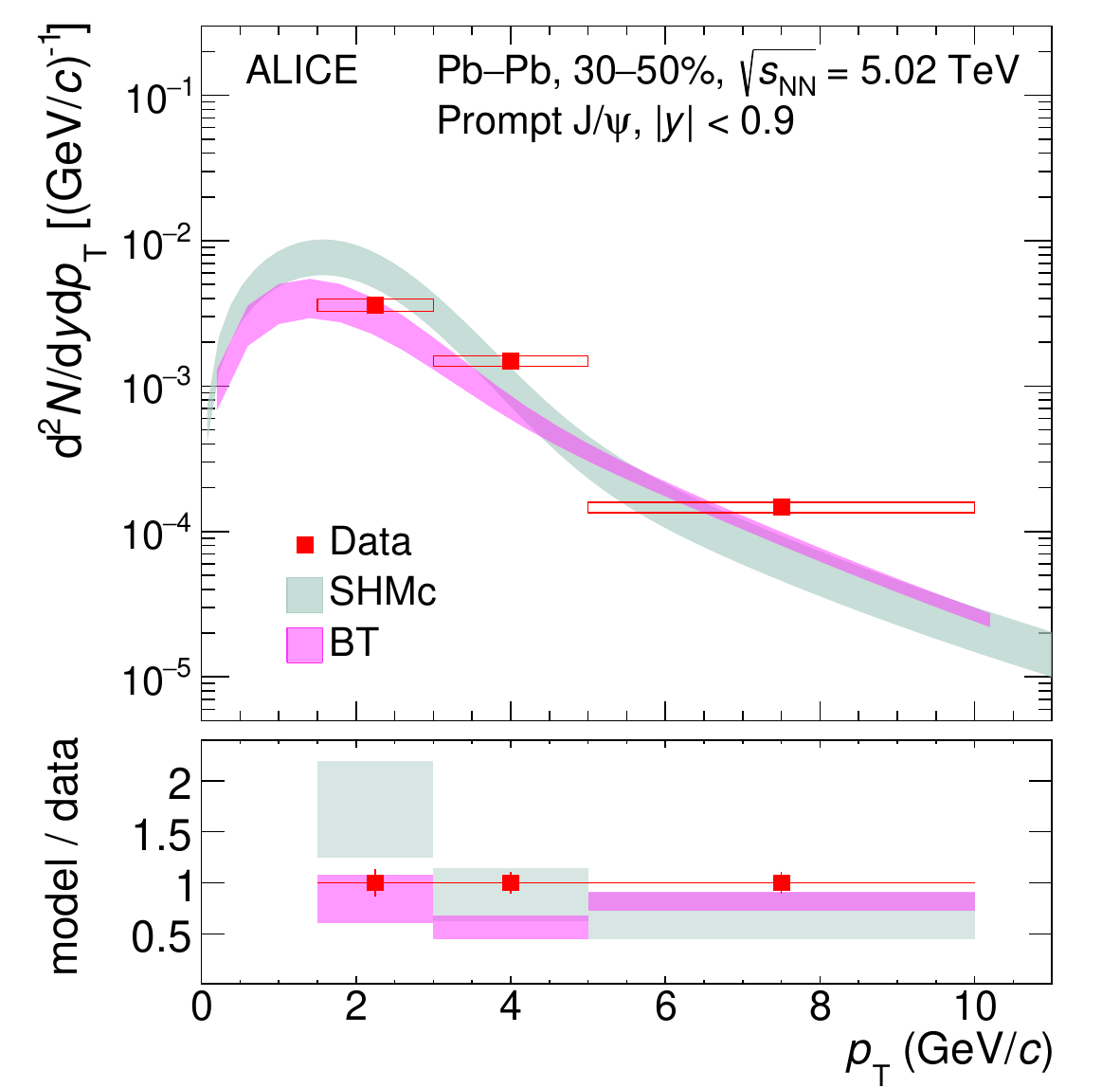}

        \caption{Prompt \jpsi yields as a function of \pt in the 0--10\% (left panel) and 30--50\% (right panel) centrality classes compared with models~\cite{Andronic:2019wva,Zhou:2014kka,Chen:2018kfo}. 
        Vertical error bars and boxes represent statistical and systematic uncertainties, respectively. Shaded bands in the top panels represent model uncertainties. Bottom panels show the ratios between models and data, with the bands representing the relative uncertainties of the models. Error bars around unity are the quadratic sum of statistical and systematic uncertainties on the measured yields.
        }
  \label{Fig:promptjpsiYieldsWithModels}
\end{figure}

Figure~\ref{Fig:promptjpsiYieldsWithModels} shows the \pt-differential yields of prompt \jpsi in the centrality classes 0--10\% (left) and 30--50\% (right), compared with
different models, namely the statistical hadronisation model (SHMc) by Andronic {\it et al.}~\cite{Andronic:2019wva} and the Boltzmann transport model (BT) by Zhuang {\it et al.}~\cite{Zhou:2014kka,Chen:2018kfo}. 
The bottom panels present the ratio of models to data, with the error bands representing the relative uncertainties originating from the models. For the computation of the ratio, an average value of the model is computed within the corresponding \pt intervals where the measurements are performed. Error bars around unity are the sum in quadrature of statistical and systematic uncertainties on the measured yields. In the SHMc model, the totality of charm quarks are
produced in the initial hard parton--parton scatterings and thermalise inside the QGP. The total charm cross section employed in the SHMc calculations is the one measured
by the ALICE collaboration in \PbPb collisions at \snn = 5.02 TeV, extracted from open-charm meson production measurements~\cite{ALICE:2022tji}. Effects from CNM are taken into account when calculating the total number of charm quarks in \PbPb using rapidity dependent measurements of the nuclear modification factor of D mesons in proton--nucleus collisions~\cite{LHCb:2017yua}, where interpolations, if necessary, are done via model calculations. The yield of the different bound states is determined by thermal weights computed at a common chemical freeze-out. 
The relative abundances of open and hidden charm hadrons are determined using the equilibrium thermodynamical parameters obtained from fits to the measured yields of light-flavoured hadrons and D-mesons, with the latter being used to extract the charm fugacity. The spectra in the SHMc model are obtained by coupling the hadron yields with a modified blast-wave function with input from hydrodynamical calculations for the flow profiles~\cite{Andronic:2019wva}.
In the BT model, the total charm cross section in \PbPb collisions is evaluated from the charm-production cross section measured in pp collisions~\cite{ALICE:2021dhb} scaled by the number of binary collisions, while to estimate CNM effects the EPS09~\cite{Eskola:2009uj} gluon distributions are used.
The dynamical evolution of the prompt charmonium states in the hot medium is described with a Boltzmann-type transport equation, including terms of dissociation
and regeneration. The dissociation of charmonia inside the medium arises from the melting of the bound states due to colour Debye screening, as well as from collisional processes of charmonia with the medium constituents, and in particular from gluon dissociation. The regeneration cross section is connected to the dissociation cross section via the detailed balance between the gluon dissociation and reversed process. The model employs the (2+1)-dimensional version of the ideal hydrodynamic equations, including both a deconfined and a hadronic phase with a first order phase transition between these two. The uncertainties plotted in Fig.~\ref{Fig:promptjpsiYieldsWithModels} for all models include contributions from total $\rm c\overline{\rm c}$ cross section, as well as uncertainties from CNM assumptions. Both SHMc and BT models show an overall good agreement with data within uncertainties, in particular for \pt below 5~\GeVc. At higher \pt, both models tend to underpredict the data, with the SHMc model showing a larger discrepancy, which is mainly due to the fact that in the SHMc model most of the produced \jpsi yields are thermal, with only a small contribution from the collision corona.
\begin{figure}[h!]
  \includegraphics[width=0.52\textwidth]{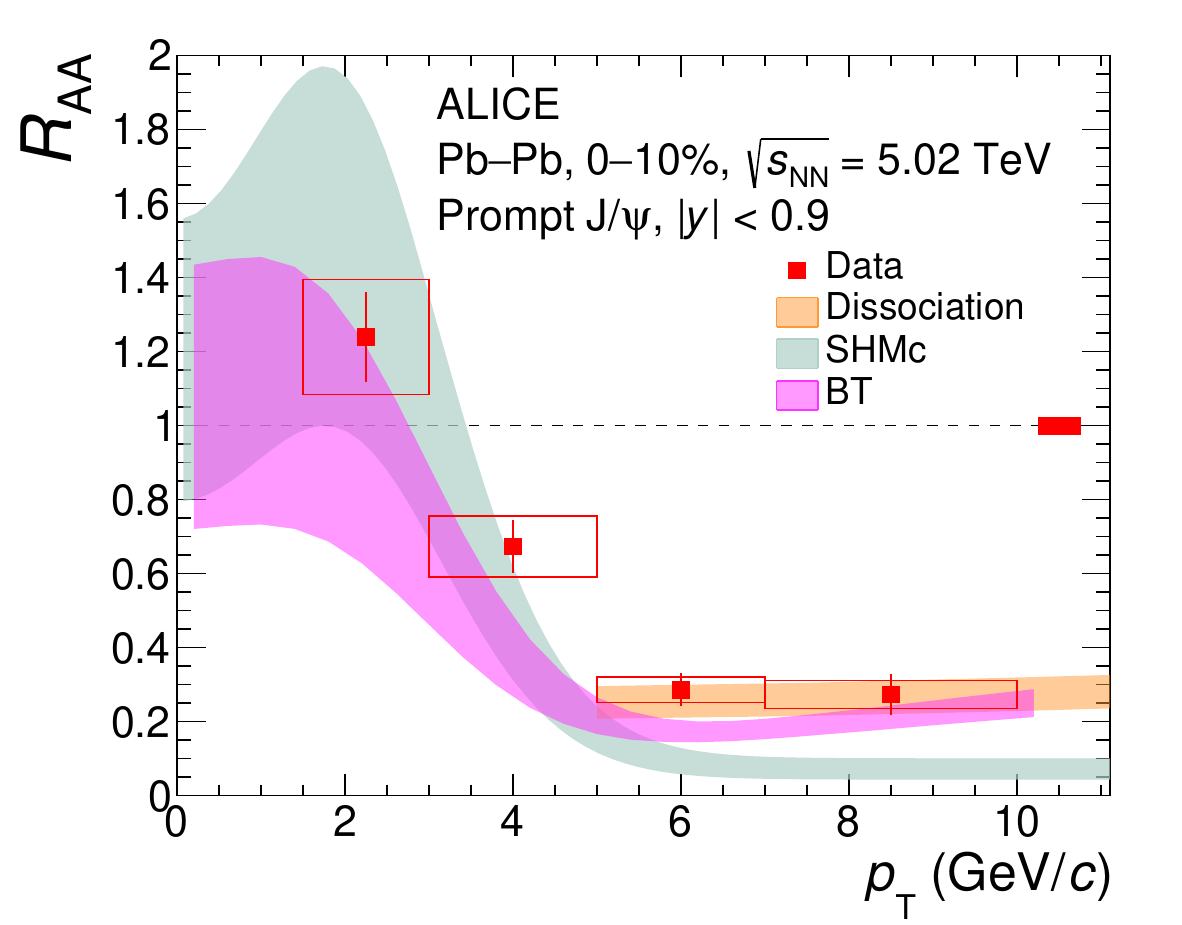}
  \includegraphics[width=0.52\textwidth]{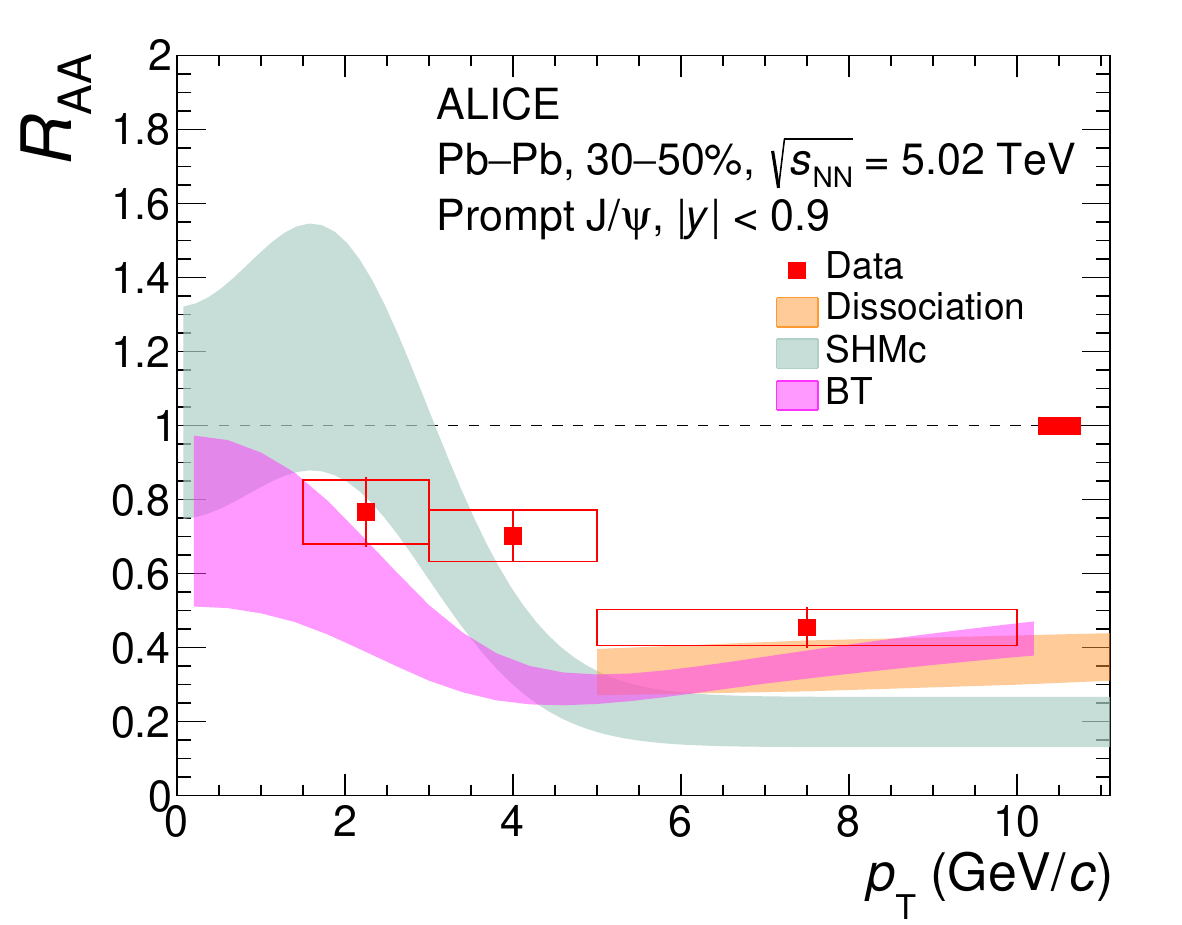}
        \caption{Prompt \jpsi \RAA as a function of \pt in the 0--10\% (left panel) and 30--50\% (right panel) centrality classes compared with models~\cite{Andronic:2019wva,Aronson:2017ymv,Makris:2019ttx,Zhou:2014kka,Chen:2018kfo}. Error bars and boxes represent statistical and uncorrelated systematic uncertainties, respectively. Shaded bands represent model uncertainties. The global uncertainty is shown around unity.
        }
  \label{Fig:promptjpsiRaaWithModels}
\end{figure}
\begin{figure}[t!]
\centering
  \includegraphics[width=0.59\textwidth]{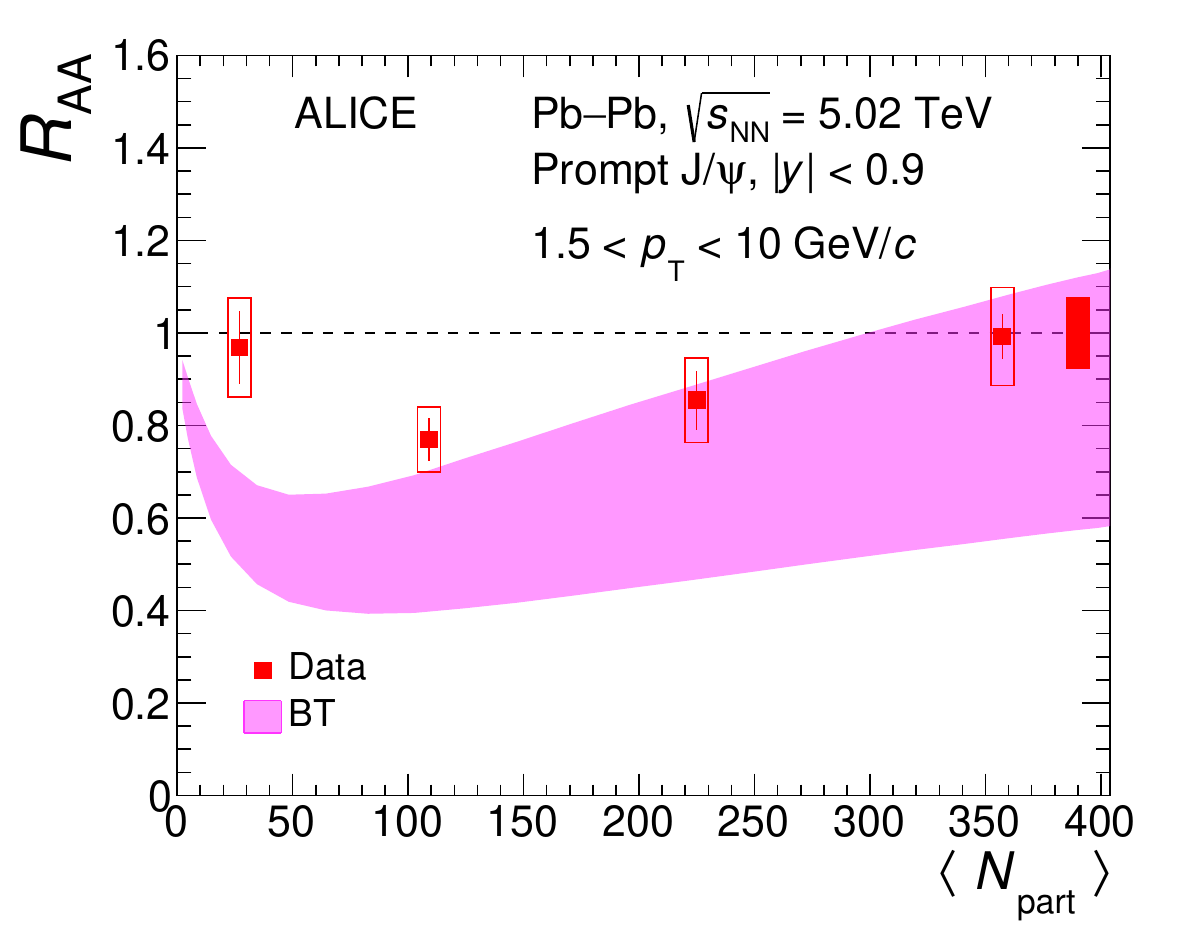}
        \caption{Centrality dependence (expressed in terms of average number of participants) of prompt \jpsi \RAA measured by ALICE in \PbPb collisions at \snn = 5.02 TeV in the transverse momentum interval 1.5 $<$ \pt $<$ 10 \GeVc. Results are compared with the BT model by Zhuang {\it et al.}~\cite{Zhou:2014kka,Chen:2018kfo}. Error bars and boxes represent statistical and uncorrelated systematic uncertainties, respectively. Shaded bands represent model uncertainties. The global uncertainty is shown around unity.
        }
  \label{Fig:promptRAAVsCentralityWithModels}
\end{figure}

Figure~\ref{Fig:promptjpsiRaaWithModels} presents the prompt \jpsi nuclear modification factor as a function of \pt in the centrality class 0--10\% (left panel) and 30--50\% (right panel) compared with model calculations. In addition to the SHMc and BT models, prompt \jpsi \RAA measurements are also compared with the dissociation model by Vitev {\it et al.}~\cite{Aronson:2017ymv}. 
In this model, which employs rate equations, the collisional dissociation of charmonia includes thermal effects on the wave function due to the screening of the ${\rm c}\overline{\rm c}$ attractive potential from the free colour charges in the QGP. The medium is modelled by a (2+1)-dimensional viscous hydrodynamic model. Non-relativistic quantum chromodynamics (NRQCD) theory~\cite{Bodwin:1994jh} is used to obtain the baseline nucleon--nucleon cross sections for charmonia and the \pt-dependent feed-down from excited states. As this model provides predictions for \pt $>$ 5~\GeVc, the contribution from CNM effects is assumed to be negligible. 
The SHMc model reproduces the prompt \jpsi \RAA results at low \pt in both centrality classes, while it is significantly below the data for \pt $>$ 5~\GeVc. The BT model provides a good description of the measurements in the full \pt range in 0--10\% most central collisions, while the model underpredicts the data in the centrality class 30--50\%. 
The dissociation model, available only above 5~\GeVc, provides a good description of prompt \jpsi \RAA measurements within uncertainties.

The centrality dependence of the \pt-integrated ($1.5 < \pt < 10$~\GeVc) \RAA of prompt \jpsi is compared with calculations from BT model in Fig.~\ref{Fig:promptRAAVsCentralityWithModels}. The BT model, which shows a rising trend with increasing number of participants from $\langle N_{\rm part} \rangle \sim 50$, is in good agreement with experimental results in 0--10\% and 10--30\% centrality classes. Below $\langle N_{\rm part} \rangle \sim 50$, both the data and the model exhibit a similar increasing trend towards more peripheral collisions, however the agreement between data and model worsens.

\subsection{Comparison with models for non-prompt \jpsi production}

In the following, the comparison of non-prompt \jpsi measurements with models describing open heavy-flavour production is discussed. As both the production mechanisms and the interaction with the medium are significantly different for prompt charmonia and open heavy-flavour hadrons, non-prompt \jpsi measurements are compared with a different set of models with respect to those considered for prompt \jpsi results.

\begin{figure}[t!]
  \includegraphics[width=0.51\textwidth]{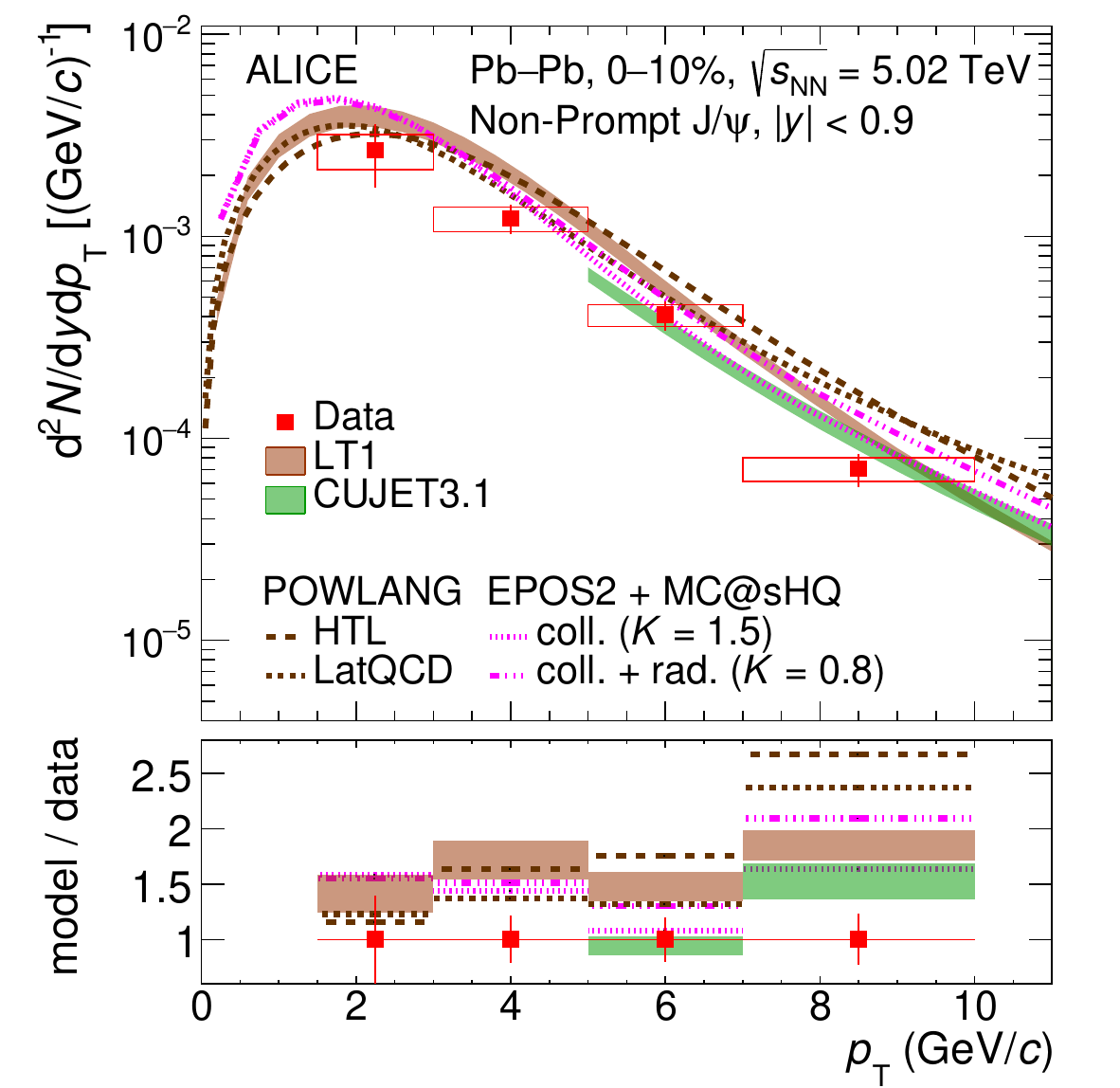}
  \includegraphics[width=0.51\textwidth]{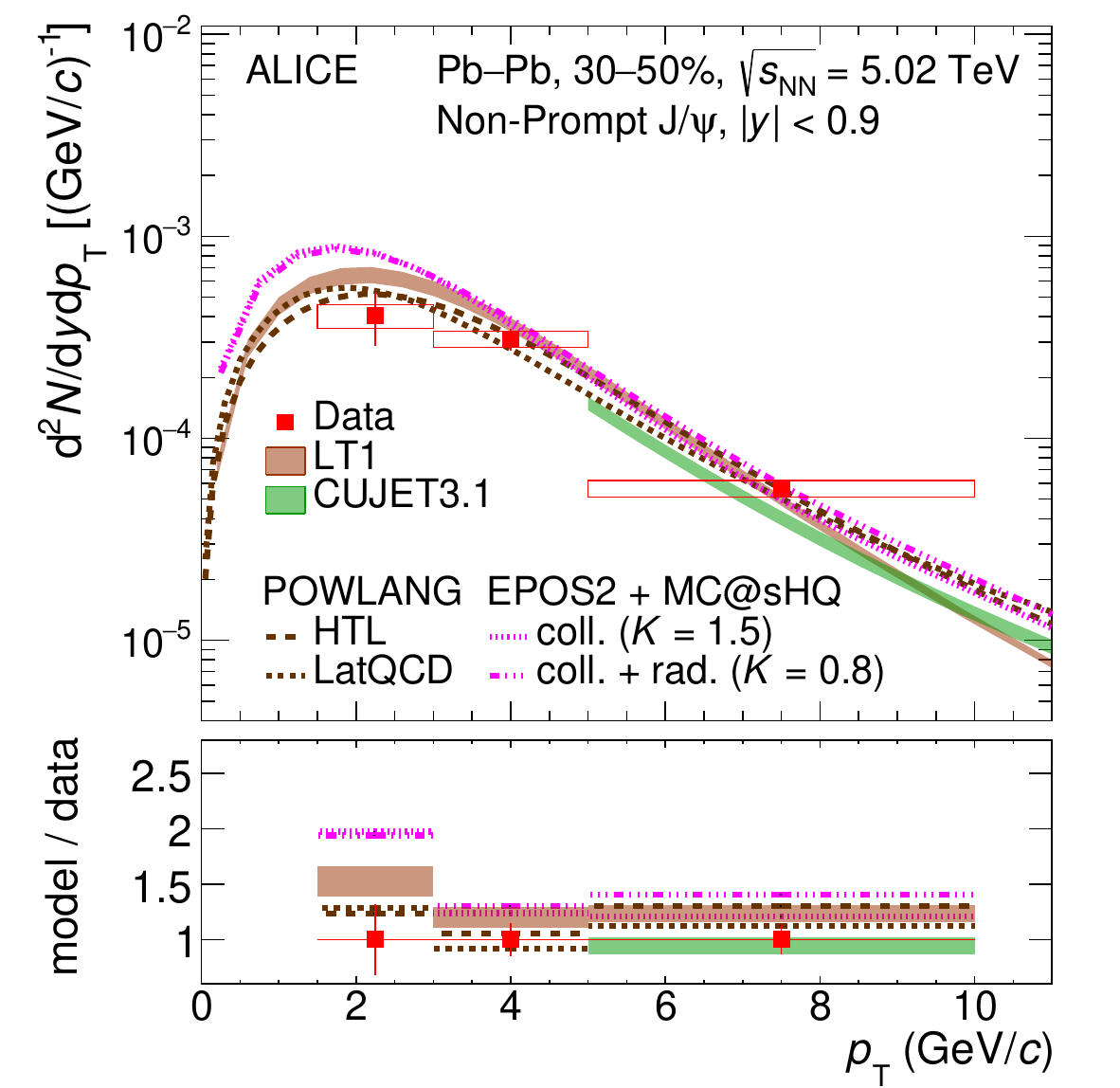}

        \caption{Non-prompt \jpsi yields as a function of \pt in the 0--10\% (left panel) and 30--50\% (right panel) centrality classes compared with models~\cite{Yang:2023rgb,Beraudo:2021ont,Beraudo:2014boa,Shi:2018izg,Shi:2018lsf,Nahrgang:2016lst}. Vertical error bars and boxes represent statistical and systematic uncertainties, respectively. Shaded bands in the top panels represent model uncertainties where applicable. Bottom panels show the ratios between models and data, with the bands representing the relative uncertainties from the models. Error bars around unity are the quadratic sum of statistical and systematic uncertainties on the measured yields.
        }
  \label{Fig:nonpromptjpsiYieldsWithModels}
\end{figure}

In Fig.~\ref{Fig:nonpromptjpsiYieldsWithModels}, the yields of non-prompt \jpsi measured in the centralities 0--10\% (left panel) and 30--50\% (right panel), are compared with partonic transport model calculations~\cite{Yang:2023rgb,Shi:2018izg,Shi:2018lsf,Beraudo:2021ont,Beraudo:2014boa,Nahrgang:2016lst}. The ratios of the models to data are depicted in the bottom panels, where the error bands are the model uncertainties. For computing the ratio, an average value of the model is considered within the corresponding \pt intervals where the measurements are performed. Error bars around unity are the quadratic sum of statistical and systematic uncertainties on the measured yields. 
In the transport model by Chen {\it et al.}~\cite{Yang:2023rgb} (LT1), as well as in the POWLANG transport model by Monteno {\it et al.}~\cite{Beraudo:2021ont,Beraudo:2014boa}, the Langevin equation is used for describing the evolution of the ancestor beauty quarks through the QGP. In POWLANG, transport coefficients are obtained either through perturbative calculations using the hard thermal loop (HTL) approach or calculations based on lattice QCD (LatQCD) simulations. The medium expansion is described by ideal (2+1)-dimensional hydrodynamic equations in LT1. In POWLANG, the background medium description is done via a (3+1)-dimensional hydrodynamic model, assuming no invariance for longitudinal boosts as considered in the (2+1)-dimensional case. The initial distribution of bottom quarks is parametrised according to perturbative-QCD calculations at fixed order with next-to-leading-log resummation (FONLL)~\cite{Cacciari:2005rk} in LT1 and with the POWHEG-BOX package~\cite{Frixione:2007nw} in POWLANG. In both models, CNM effects are accounted for by using the  EPS09~\cite{Eskola:2009uj} gluon parton distribution functions, and at the hadronisation hypersurface, bottom and light-flavour quarks hadronise into beauty mesons via the coalescence model. In LT1, medium-induced gluon radiation is also included in the energy loss of heavy quarks, and becomes the dominant mechanism at large momentum, while in POWLANG the interactions of the heavy quarks with the medium constituents happen solely via collisional processes.
In the calculations by Shi {\it et al.}~\cite{Shi:2018izg,Shi:2018lsf}, the CUJET3.1 framework is used to evaluate the jet energy loss in a (2+1)-dimensional hydrodynamic background, implementing the contributions from collisional as well as from radiative processes. A set of \RAA and elliptic flow results from light hadrons in central and semicentral heavy-ion collisions is used to constrain the 
model, which is then used to predict heavy-flavour observables. For the initial \pt distribution of beauty quarks, FONLL calculations~\cite{Cacciari:2005rk} with CTEQ6M~\cite{Pumplin:2002vw} parton distribution functions are used. The formation of beauty hadrons happens via the classical vacuum-like fragmentation using the Peterson parametrisation in Ref.~\cite{PhysRevD.27.105}. In the model by Gossiaux {\it et al.}~\cite{Nahrgang:2016lst} (EPOS2+MC@sHQ), the Monte Carlo treatment of the Boltzmann equation of heavy quarks (MC@sHQ)~\cite{Gossiaux:2008jv} is coupled to a (3+1)-dimensional fluid dynamical evolution of the locally thermalized QGP following the initial conditions from EPOS2~\cite{PhysRevC.82.044904,PhysRevC.85.064907}. The initial transverse momentum spectra of beauty quarks is obtained from FONLL calculations~\cite{Cacciari:2005rk}, and nuclear shadowing has been included according to the EPS09~\cite{Eskola:2009uj} parametrization of the nuclear parton distribution functions. 
The calculations shown in this article consider two different configurations, either including pure collisional processes or both collisional and radiative ones. 
In order to further constrain the model, the corresponding cross sections are rescaled by a global factor $K$, which is chosen such that the predictions give a reasonable agreement at intermediate and high \pt with a set of D-meson \RAA measurements available in \PbPb collisions at \snn = 2.76 TeV~\cite{Nahrgang:2016lst}. In particular, $K$ = 0.8 (1.5) is considered when both collisional and radiative (pure collisional) processes are included.  
After the evolution in the medium, the beauty quarks hadronize via both fragmentation and coalescence. The LT1, POWLANG and EPOS2+MC@sHQ models show systematically higher values compared to data, for both centrality classes in the full measured transverse momentum range. The discrepancy looks larger for POWLANG, especially in centrality class 0--10\% and at higher \pt, which could be related to the absence of radiative processes in this model. The CUJET3.1 model, available only for \pt above 5~\GeVc, is compatible with the data within uncertainties. 

\begin{figure}[t!]
  \includegraphics[width=0.52\textwidth]{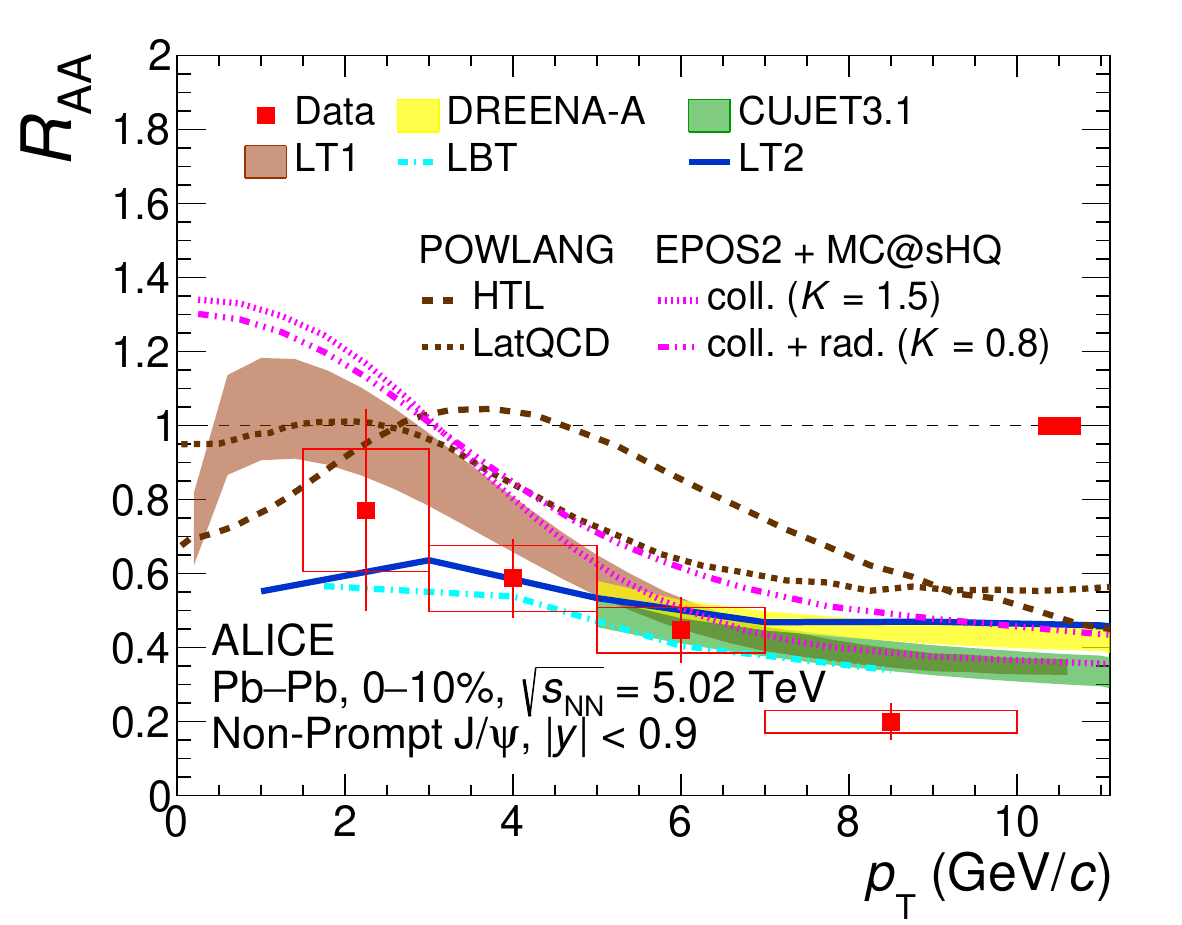}
  \includegraphics[width=0.52\textwidth]{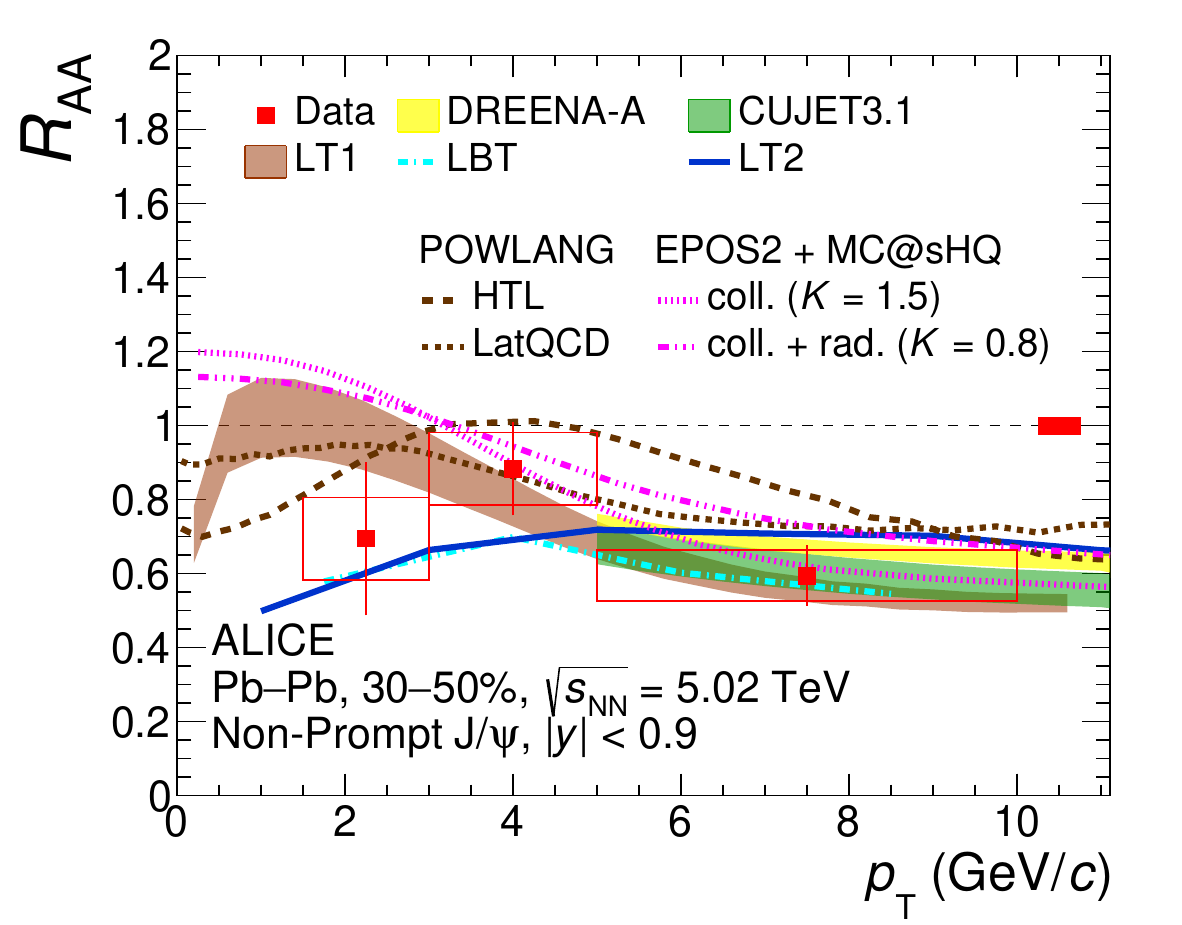}
        \caption{Non-prompt \jpsi \RAA as a function of \pt in the 0--10\% (left panel) and 30--50\% (right panel) centrality classes compared with models~\cite{Yang:2023rgb, Li:2021xbd,Xing:2021xwc,Shi:2018izg,Shi:2018lsf,Zigic:2021rku,Stojku:2020wkh,Nahrgang:2016lst}. Error bars and boxes represent statistical and uncorrelated systematic uncertainties, respectively. Shaded bands represent model uncertainties where applicable. The global uncertainty is shown around unity.
        }
  \label{Fig:nonpromptjpsiRaaWithModels}
\end{figure}

\begin{figure}[h!]
\centering
  \includegraphics[width=0.59\textwidth]{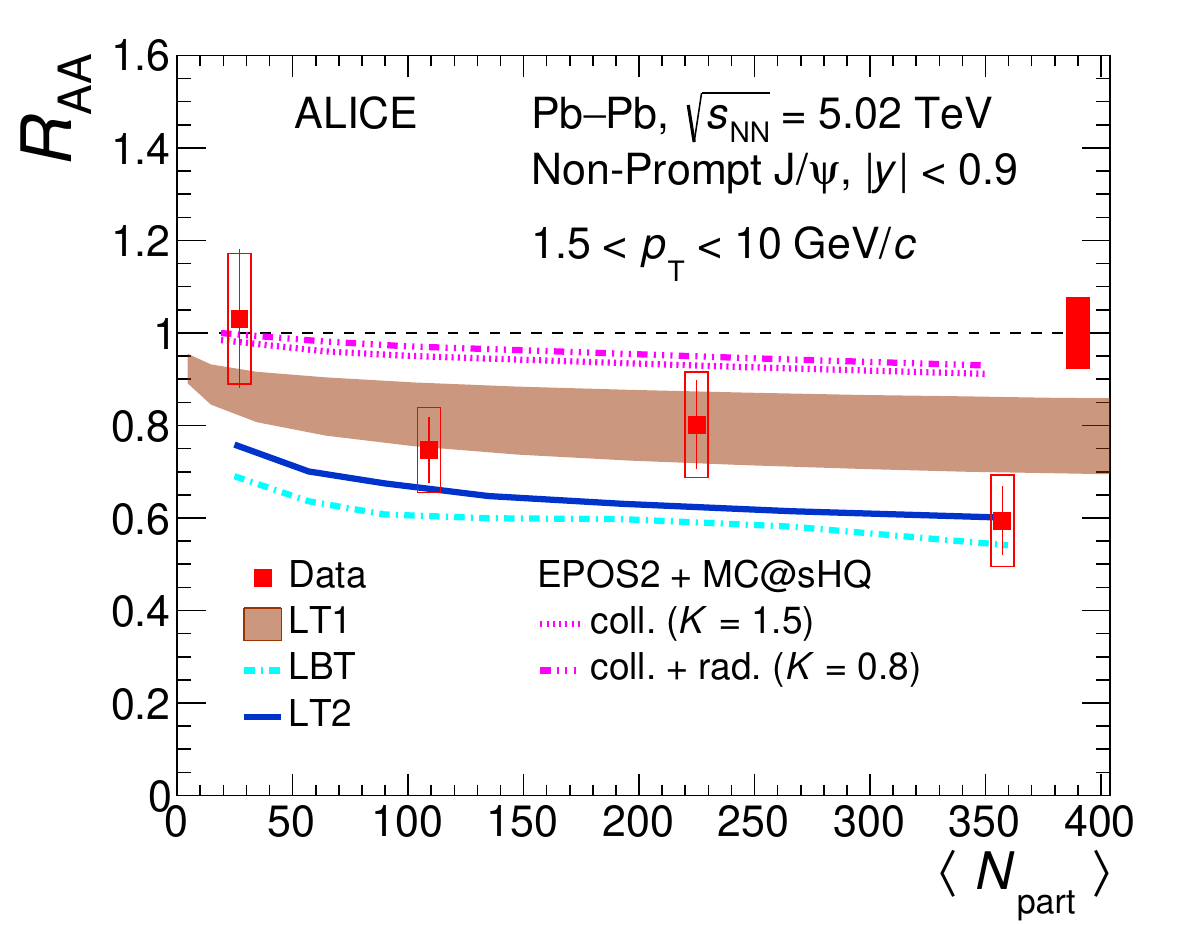}
        \caption{Centrality dependence (expressed in terms of average number of participants) of non-prompt \jpsi \RAA measured by ALICE in \PbPb collisions at \snn = 5.02 TeV in the transverse momentum interval 1.5 $<$ \pt $<$ 10 \GeVc. Results are compared with several partonic transport models~\cite{Yang:2023rgb,Li:2021xbd,Xing:2021xwc,Nahrgang:2016lst}. Error bars and boxes represent statistical and uncorrelated systematic uncertainties, respectively. Shaded bands represent model uncertainties where applicable. The global uncertainty is shown around unity
        }
  \label{Fig:nonpromptRAAVsCentralityWithModels}
\end{figure}

The nuclear modification factor of non-prompt \jpsi is compared with models in Fig.~\ref{Fig:nonpromptjpsiRaaWithModels} in 0--10\% (left panel) and 30--50\% (right panel) centrality intervals. The results are compared with the LT1, CUJET3.1, POWLANG and EPOS2+MC@sHQ models previously described as well as with additional ones~\cite{Li:2021xbd,Xing:2021xwc,Zigic:2021rku,Stojku:2020wkh}. 
 In these additional calculations, the space--time evolution of the QGP is simulated using a (3+1)-dimensional viscous hydrodynamic model and  both collisional and radiative energy loss mechanisms inside a thermal medium are considered. In the calculation by Li {\it et al.}~\cite{Li:2021xbd}, marked as LT2, 
the interactions between heavy quarks and the QGP are described by an improved Langevin approach. 
The transport model by Xing {\it et al.}~\cite{Xing:2021xwc} employs an extended linear Boltzmann transport (LBT) equation, which includes both short and long-range interactions of heavy quarks with the QGP.
In both LT2 and LBT models, the initial heavy quark \pt distribution is simulated according to FONLL calculations using CT14NLO~\cite{Dulat:2015mca} parton distribution functions modified according to the EPPS16~\cite{Eskola:2016oht} next-to-leading-order parametrisation, while beauty hadrons are produced through a
hybrid fragmentation--coalescence model~\cite{Cao:2019iqs}. 
The calculation by Djordjevic {\it et al.}~\cite{Zigic:2021rku,Stojku:2020wkh} employs a  framework (DREENA-A) which combines the 
state-of-the-art dynamical energy loss model with hydrodynamical simulations. The initial heavy-quark spectrum is computed using next-to-leading-order calculations described in Ref.~\cite{PhysRevC.80.054902} and KLP~\cite{Kartvelishvili:1977pi} fragmentation functions are used for the formation of beauty mesons. 
In both centralities all available model predictions except POWLANG show compatible values for \pt above 5~\GeVc, and within uncertainties are in an overall good agreement with the data. 
The POWLANG model overpredicts the \RAA in the centrality class 0--10\% and at high \pt, which might be a consequence of the lack of radiative energy loss contributions in this model. Below 5~\GeVc, the LT1 and POWLANG models sit on the upper side of the data points in both centrality classes, still being compatible with them within uncertainties, while the EPOS2+MC@sHQ model overpredicts the measurements. Both LBT and LT2 models are compatible with the measured \RAA within uncertainties in the full measured \pt range, and in both centrality classes.   

Figure~\ref{Fig:nonpromptRAAVsCentralityWithModels} depicts the non-prompt \jpsi \RAA, integrated over \pt in the interval 1.5 $<$ \pt $<$ 10~\GeVc, as a function of the number of participants in comparison with transport models. The LT1 model shows a slightly decreasing trend moving towards central collisions, and is compatible with data within uncertainties for all centrality classes. The LBT, LT2 and EPOS2+MC@sHQ models predict a similar decreasing trend for the non-prompt \jpsi \RAA towards larger $\langle N_{\rm part} \rangle$. The LBT and LT2 models show good agreement with \RAA results in 0--10\% centrality class, while for other centrality classes both models slightly underpredict the data. The EPOS2+MC@sHQ model agrees with measured \RAA in the most peripheral class, while it tends to overestimate measurements towards more central collisions.  

\section{Summary}
\label{Sec:summary}

Prompt and non-prompt \jpsi production is measured by the ALICE collaboration in \PbPb collisions at \snn = 5.02 TeV in the rapidity interval $|y| < 0.9$ as a function of \pt and centrality. In particular, \pt-differential measurements of non-prompt \jpsi fractions, production yields, and nuclear modification factors are carried out. 
The ALICE results extend the existing CMS and ATLAS measurements at midrapidity, available only at high \pt, down to \pt = 1.5 \GeVc, and all measurements look compatible within uncertainties in the common \pt intervals.
 Non-prompt \jpsi fractions show a rising trend with increasing \pt, similar to the one observed in pp collisions, while no significant dependence on the centrality is observed within uncertainties, with the exception of most central collisions where \fb exhibits a significant decrease with respect to other centrality classes. The comparison with earlier measurements in Pb--Pb collisions at \snn = 2.76 TeV shows compatible results at the two centre-of-mass energies and highlights a significantly improved precision with Run 2 data. 

For \pt $>$ 5 \GeVc, the prompt \jpsi \RAA decreases with increasing centrality, while at lower \pt the suppression is smaller in 0--10\% most central collisions, in particular in the lowest \pt interval where the prompt \jpsi \RAA exceeds unity. These results are consistent with \pt-integrated measurements of prompt \jpsi \RAA, which rises with  $\langle N_{\rm part} \rangle$, hinting at an increasing contribution from regeneration at low \pt and more central collisions.
In 0--10\% most central collisions, the SHMc model and transport microscopic calculations that include a contribution from regeneration are compatible with experimental data for $\pt < 5$~\GeVc. However, at higher \pt, transport models are compatible with the measurements within uncertainties, while the SHMc model significantly underpredicts the data. In semicentral collisions there is less agreement between data and models. In particular, for \pt above 3 \GeVc the models either underpredict the data or sit at the lower edge of the experimental uncertainty.

The non-prompt \jpsi \RAA integrated over \pt hints at a decreasing trend towards more central collisions. As a function of \pt, the non-prompt \jpsi \RAA in different centrality classes are compatible within uncertainties below 5 \GeVc, while at higher \pt the suppression is larger in the centrality interval 0--10\%. Results are consistent within uncertainties with non-prompt ${\rm D}^{\rm  0}$ \RAA measurements in the centrality classes 0--10\% and 30--50\%. Several transport models are able to describe the data within uncertainties. All calculations, but POWLANG, implement both collisional and radiative energy loss processes combined with a dynamically expanding QGP, considering different hypotheses on transport dynamics, CNM effects, \pt distributions and hadronisation of beauty quarks. Above 5~\GeVc, all calculations predict a similar suppression for non-prompt \jpsi and are consistent with the data within uncertainties, with the exception of POWLANG, which overpredicts the data in most central events. This points to the importance of radiative energy loss contributions in the high-\pt region. At lower \pt and for the \pt-integrated case, theoretical calculations predict different magnitudes of the suppression. However, due to the current precision of the measurements it is not possible to discriminate among them as all calculations are compatible with the data within uncertainties. 

Both prompt and non-prompt \jpsi measurements will improve significantly taking advantage of the larger expected data sample and the better spatial resolution provided at midrapidity by the upgraded ITS~\cite{ALICE:2023udb} in the next LHC runs. 
In particular, further differential measurements in the non-prompt charmonium sector, in addition to yet unmeasured observables such as the elliptic flow, could allow further constraining different open beauty hadron production models.


\newenvironment{acknowledgement}{\relax}{\relax}
\begin{acknowledgement}
\section*{Acknowledgements}

The ALICE Collaboration would like to thank all its engineers and technicians for their invaluable contributions to the construction of the experiment and the CERN accelerator teams for the outstanding performance of the LHC complex.
The ALICE Collaboration gratefully acknowledges the resources and support provided by all Grid centres and the Worldwide LHC Computing Grid (WLCG) collaboration.
The ALICE Collaboration acknowledges the following funding agencies for their support in building and running the ALICE detector:
A. I. Alikhanyan National Science Laboratory (Yerevan Physics Institute) Foundation (ANSL), State Committee of Science and World Federation of Scientists (WFS), Armenia;
Austrian Academy of Sciences, Austrian Science Fund (FWF): [M 2467-N36] and Nationalstiftung f\"{u}r Forschung, Technologie und Entwicklung, Austria;
Ministry of Communications and High Technologies, National Nuclear Research Center, Azerbaijan;
Conselho Nacional de Desenvolvimento Cient\'{\i}fico e Tecnol\'{o}gico (CNPq), Financiadora de Estudos e Projetos (Finep), Funda\c{c}\~{a}o de Amparo \`{a} Pesquisa do Estado de S\~{a}o Paulo (FAPESP) and Universidade Federal do Rio Grande do Sul (UFRGS), Brazil;
Bulgarian Ministry of Education and Science, within the National Roadmap for Research Infrastructures 2020-2027 (object CERN), Bulgaria;
Ministry of Education of China (MOEC) , Ministry of Science \& Technology of China (MSTC) and National Natural Science Foundation of China (NSFC), China;
Ministry of Science and Education and Croatian Science Foundation, Croatia;
Centro de Aplicaciones Tecnol\'{o}gicas y Desarrollo Nuclear (CEADEN), Cubaenerg\'{\i}a, Cuba;
Ministry of Education, Youth and Sports of the Czech Republic, Czech Republic;
The Danish Council for Independent Research | Natural Sciences, the VILLUM FONDEN and Danish National Research Foundation (DNRF), Denmark;
Helsinki Institute of Physics (HIP), Finland;
Commissariat \`{a} l'Energie Atomique (CEA) and Institut National de Physique Nucl\'{e}aire et de Physique des Particules (IN2P3) and Centre National de la Recherche Scientifique (CNRS), France;
Bundesministerium f\"{u}r Bildung und Forschung (BMBF) and GSI Helmholtzzentrum f\"{u}r Schwerionenforschung GmbH, Germany;
General Secretariat for Research and Technology, Ministry of Education, Research and Religions, Greece;
National Research, Development and Innovation Office, Hungary;
Department of Atomic Energy Government of India (DAE), Department of Science and Technology, Government of India (DST), University Grants Commission, Government of India (UGC) and Council of Scientific and Industrial Research (CSIR), India;
National Research and Innovation Agency - BRIN, Indonesia;
Istituto Nazionale di Fisica Nucleare (INFN), Italy;
Japanese Ministry of Education, Culture, Sports, Science and Technology (MEXT) and Japan Society for the Promotion of Science (JSPS) KAKENHI, Japan;
Consejo Nacional de Ciencia (CONACYT) y Tecnolog\'{i}a, through Fondo de Cooperaci\'{o}n Internacional en Ciencia y Tecnolog\'{i}a (FONCICYT) and Direcci\'{o}n General de Asuntos del Personal Academico (DGAPA), Mexico;
Nederlandse Organisatie voor Wetenschappelijk Onderzoek (NWO), Netherlands;
The Research Council of Norway, Norway;
Commission on Science and Technology for Sustainable Development in the South (COMSATS), Pakistan;
Pontificia Universidad Cat\'{o}lica del Per\'{u}, Peru;
Ministry of Education and Science, National Science Centre and WUT ID-UB, Poland;
Korea Institute of Science and Technology Information and National Research Foundation of Korea (NRF), Republic of Korea;
Ministry of Education and Scientific Research, Institute of Atomic Physics, Ministry of Research and Innovation and Institute of Atomic Physics and University Politehnica of Bucharest, Romania;
Ministry of Education, Science, Research and Sport of the Slovak Republic, Slovakia;
National Research Foundation of South Africa, South Africa;
Swedish Research Council (VR) and Knut \& Alice Wallenberg Foundation (KAW), Sweden;
European Organization for Nuclear Research, Switzerland;
Suranaree University of Technology (SUT), National Science and Technology Development Agency (NSTDA), Thailand Science Research and Innovation (TSRI) and National Science, Research and Innovation Fund (NSRF), Thailand;
Turkish Energy, Nuclear and Mineral Research Agency (TENMAK), Turkey;
National Academy of  Sciences of Ukraine, Ukraine;
Science and Technology Facilities Council (STFC), United Kingdom;
National Science Foundation of the United States of America (NSF) and United States Department of Energy, Office of Nuclear Physics (DOE NP), United States of America.
In addition, individual groups or members have received support from:
European Research Council, Strong 2020 - Horizon 2020 (grant nos. 950692, 824093), European Union;
Academy of Finland (Center of Excellence in Quark Matter) (grant nos. 346327, 346328), Finland.

\end{acknowledgement}

\bibliographystyle{utphys}   
\bibliography{bibliography}

\newpage
\appendix

%
%

\section{The ALICE Collaboration}
\label{app:collab}
\begin{flushleft} 
\small

S.~Acharya\,\orcidlink{0000-0002-9213-5329}\,$^{\rm 128}$, 
D.~Adamov\'{a}\,\orcidlink{0000-0002-0504-7428}\,$^{\rm 87}$, 
G.~Aglieri Rinella\,\orcidlink{0000-0002-9611-3696}\,$^{\rm 33}$, 
M.~Agnello\,\orcidlink{0000-0002-0760-5075}\,$^{\rm 30}$, 
N.~Agrawal\,\orcidlink{0000-0003-0348-9836}\,$^{\rm 52}$, 
Z.~Ahammed\,\orcidlink{0000-0001-5241-7412}\,$^{\rm 136}$, 
S.~Ahmad\,\orcidlink{0000-0003-0497-5705}\,$^{\rm 16}$, 
S.U.~Ahn\,\orcidlink{0000-0001-8847-489X}\,$^{\rm 72}$, 
I.~Ahuja\,\orcidlink{0000-0002-4417-1392}\,$^{\rm 38}$, 
A.~Akindinov\,\orcidlink{0000-0002-7388-3022}\,$^{\rm 142}$
% ...

\section*{Affiliation Notes}

$^{\rm I}$ Deceased\\
$^{\rm II}$ Also at: Max-Planck-Institut fur Physik, Munich, Germany\\
$^{\rm III}$ Also at: Italian National Agency for New Technologies, Energy and Sustainable Economic Development (ENEA), Bologna, Italy\\
$^{\rm IV}$ Also at: Dipartimento DET del Politecnico di Torino, Turin, Italy\\
$^{\rm V}$ Also at: Department of Applied Physics, Aligarh Muslim University, Aligarh, India\\
$^{\rm VI}$ Also at: Institute of Theoretical Physics, University of Wroclaw, Poland\\
$^{\rm VII}$ Also at: An institution covered by a cooperation agreement with CERN\\

\section*{Collaboration Institutes}

$^{1}$ A.I. Alikhanyan National Science Laboratory (Yerevan Physics Institute) Foundation, Yerevan, Armenia\\
$^{2}$ AGH University of Krakow, Cracow, Poland\\
$^{3}$ Bogolyubov Institute for Theoretical Physics, National Academy of Sciences of Ukraine, Kiev, Ukraine\\
$^{4}$ Bose Institute, Department of Physics  and Centre for Astroparticle Physics and Space Science (CAPSS), Kolkata, India\\
$^{5}$ California Polytechnic State University, San Luis Obispo, California, United States\\
$^{6}$ Central China Normal University, Wuhan, China\\
$^{7}$ Centro de Aplicaciones Tecnol\'{o}gicas y Desarrollo Nuclear (CEADEN), Havana, Cuba\\
$^{8}$ Centro de Investigaci\'{o}n y de Estudios Avanzados (CINVESTAV), Mexico City and M\'{e}rida, Mexico\\
$^{9}$ Chicago State University, Chicago, Illinois, United States\\
$^{10}$ China Institute of Atomic Energy, Beijing, China\\
$^{11}$ China University of Geosciences, Wuhan, China\\
$^{12}$ Chungbuk National University, Cheongju, Republic of Korea\\
$^{13}$ Comenius University Bratislava, Faculty of Mathematics, Physics and Informatics, Bratislava, Slovak Republic\\
$^{14}$ COMSATS University Islamabad, Islamabad, Pakistan\\
$^{15}$ Creighton University, Omaha, Nebraska, United States\\
$^{16}$ Department of Physics, Aligarh Muslim University, Aligarh, India\\
$^{17}$ Department of Physics, Pusan National University, Pusan, Republic of Korea\\
$^{18}$ Department of Physics, Sejong University, Seoul, Republic of Korea\\
$^{19}$ Department of Physics, University of California, Berkeley, California, United States\\
$^{20}$ Department of Physics, University of Oslo, Oslo, Norway\\
$^{21}$ Department of Physics and Technology, University of Bergen, Bergen, Norway\\
$^{22}$ Dipartimento di Fisica, Universit\`{a} di Pavia, Pavia, Italy\\
$^{23}$ Dipartimento di Fisica dell'Universit\`{a} and Sezione INFN, Cagliari, Italy\\
$^{24}$ Dipartimento di Fisica dell'Universit\`{a} and Sezione INFN, Trieste, Italy\\
$^{25}$ Dipartimento di Fisica dell'Universit\`{a} and Sezione INFN, Turin, Italy\\
$^{26}$ Dipartimento di Fisica e Astronomia dell'Universit\`{a} and Sezione INFN, Bologna, Italy\\
$^{27}$ Dipartimento di Fisica e Astronomia dell'Universit\`{a} and Sezione INFN, Catania, Italy\\
$^{28}$ Dipartimento di Fisica e Astronomia dell'Universit\`{a} and Sezione INFN, Padova, Italy\\
$^{29}$ Dipartimento di Fisica `E.R.~Caianiello' dell'Universit\`{a} and Gruppo Collegato INFN, Salerno, Italy\\
$^{30}$ Dipartimento DISAT del Politecnico and Sezione INFN, Turin, Italy\\
$^{31}$ Dipartimento di Scienze MIFT, Universit\`{a} di Messina, Messina, Italy\\
$^{32}$ Dipartimento Interateneo di Fisica `M.~Merlin' and Sezione INFN, Bari, Italy\\
$^{33}$ European Organization for Nuclear Research (CERN), Geneva, Switzerland\\
$^{34}$ Faculty of Electrical Engineering, Mechanical Engineering and Naval Architecture, University of Split, Split, Croatia\\
$^{35}$ Faculty of Engineering and Science, Western Norway University of Applied Sciences, Bergen, Norway\\
$^{36}$ Faculty of Nuclear Sciences and Physical Engineering, Czech Technical University in Prague, Prague, Czech Republic\\
$^{37}$ Faculty of Physics, Sofia University, Sofia, Bulgaria\\
$^{38}$ Faculty of Science, P.J.~\v{S}af\'{a}rik University, Ko\v{s}ice, Slovak Republic\\
$^{39}$ Frankfurt Institute for Advanced Studies, Johann Wolfgang Goethe-Universit\"{a}t Frankfurt, Frankfurt, Germany\\
$^{40}$ Fudan University, Shanghai, China\\
$^{41}$ Gangneung-Wonju National University, Gangneung, Republic of Korea\\
$^{42}$ Gauhati University, Department of Physics, Guwahati, India\\
$^{43}$ Helmholtz-Institut f\"{u}r Strahlen- und Kernphysik, Rheinische Friedrich-Wilhelms-Universit\"{a}t Bonn, Bonn, Germany\\
$^{44}$ Helsinki Institute of Physics (HIP), Helsinki, Finland\\
$^{45}$ High Energy Physics Group,  Universidad Aut\'{o}noma de Puebla, Puebla, Mexico\\
$^{46}$ Horia Hulubei National Institute of Physics and Nuclear Engineering, Bucharest, Romania\\
$^{47}$ HUN-REN Wigner Research Centre for Physics, Budapest, Hungary\\
$^{48}$ Indian Institute of Technology Bombay (IIT), Mumbai, India\\
$^{49}$ Indian Institute of Technology Indore, Indore, India\\
$^{50}$ INFN, Laboratori Nazionali di Frascati, Frascati, Italy\\
$^{51}$ INFN, Sezione di Bari, Bari, Italy\\
$^{52}$ INFN, Sezione di Bologna, Bologna, Italy\\
$^{53}$ INFN, Sezione di Cagliari, Cagliari, Italy\\
$^{54}$ INFN, Sezione di Catania, Catania, Italy\\
$^{55}$ INFN, Sezione di Padova, Padova, Italy\\
$^{56}$ INFN, Sezione di Pavia, Pavia, Italy\\
$^{57}$ INFN, Sezione di Torino, Turin, Italy\\
$^{58}$ INFN, Sezione di Trieste, Trieste, Italy\\
$^{59}$ Inha University, Incheon, Republic of Korea\\
$^{60}$ Institute for Gravitational and Subatomic Physics (GRASP), Utrecht University/Nikhef, Utrecht, Netherlands\\
$^{61}$ Institute of Experimental Physics, Slovak Academy of Sciences, Ko\v{s}ice, Slovak Republic\\
$^{62}$ Institute of Physics, Homi Bhabha National Institute, Bhubaneswar, India\\
$^{63}$ Institute of Physics of the Czech Academy of Sciences, Prague, Czech Republic\\
$^{64}$ Institute of Space Science (ISS), Bucharest, Romania\\
$^{65}$ Institut f\"{u}r Kernphysik, Johann Wolfgang Goethe-Universit\"{a}t Frankfurt, Frankfurt, Germany\\
$^{66}$ Instituto de Ciencias Nucleares, Universidad Nacional Aut\'{o}noma de M\'{e}xico, Mexico City, Mexico\\
$^{67}$ Instituto de F\'{i}sica, Universidade Federal do Rio Grande do Sul (UFRGS), Porto Alegre, Brazil\\
$^{68}$ Instituto de F\'{\i}sica, Universidad Nacional Aut\'{o}noma de M\'{e}xico, Mexico City, Mexico\\
$^{69}$ iThemba LABS, National Research Foundation, Somerset West, South Africa\\
$^{70}$ Jeonbuk National University, Jeonju, Republic of Korea\\
$^{71}$ Johann-Wolfgang-Goethe Universit\"{a}t Frankfurt Institut f\"{u}r Informatik, Fachbereich Informatik und Mathematik, Frankfurt, Germany\\
$^{72}$ Korea Institute of Science and Technology Information, Daejeon, Republic of Korea\\
$^{73}$ KTO Karatay University, Konya, Turkey\\
$^{74}$ Laboratoire de Physique Subatomique et de Cosmologie, Universit\'{e} Grenoble-Alpes, CNRS-IN2P3, Grenoble, France\\
$^{75}$ Lawrence Berkeley National Laboratory, Berkeley, California, United States\\
$^{76}$ Lund University Department of Physics, Division of Particle Physics, Lund, Sweden\\
$^{77}$ Nagasaki Institute of Applied Science, Nagasaki, Japan\\
$^{78}$ Nara Women{'}s University (NWU), Nara, Japan\\
$^{79}$ National and Kapodistrian University of Athens, School of Science, Department of Physics , Athens, Greece\\
$^{80}$ National Centre for Nuclear Research, Warsaw, Poland\\
$^{81}$ National Institute of Science Education and Research, Homi Bhabha National Institute, Jatni, India\\
$^{82}$ National Nuclear Research Center, Baku, Azerbaijan\\
$^{83}$ National Research and Innovation Agency - BRIN, Jakarta, Indonesia\\
$^{84}$ Niels Bohr Institute, University of Copenhagen, Copenhagen, Denmark\\
$^{85}$ Nikhef, National institute for subatomic physics, Amsterdam, Netherlands\\
$^{86}$ Nuclear Physics Group, STFC Daresbury Laboratory, Daresbury, United Kingdom\\
$^{87}$ Nuclear Physics Institute of the Czech Academy of Sciences, Husinec-\v{R}e\v{z}, Czech Republic\\
$^{88}$ Oak Ridge National Laboratory, Oak Ridge, Tennessee, United States\\
$^{89}$ Ohio State University, Columbus, Ohio, United States\\
$^{90}$ Physics department, Faculty of science, University of Zagreb, Zagreb, Croatia\\
$^{91}$ Physics Department, Panjab University, Chandigarh, India\\
$^{92}$ Physics Department, University of Jammu, Jammu, India\\
$^{93}$ Physics Program and International Institute for Sustainability with Knotted Chiral Meta Matter (SKCM2), Hiroshima University, Hiroshima, Japan\\
$^{94}$ Physikalisches Institut, Eberhard-Karls-Universit\"{a}t T\"{u}bingen, T\"{u}bingen, Germany\\
$^{95}$ Physikalisches Institut, Ruprecht-Karls-Universit\"{a}t Heidelberg, Heidelberg, Germany\\
$^{96}$ Physik Department, Technische Universit\"{a}t M\"{u}nchen, Munich, Germany\\
$^{97}$ Politecnico di Bari and Sezione INFN, Bari, Italy\\
$^{98}$ Research Division and ExtreMe Matter Institute EMMI, GSI Helmholtzzentrum f\"ur Schwerionenforschung GmbH, Darmstadt, Germany\\
$^{99}$ Saga University, Saga, Japan\\
$^{100}$ Saha Institute of Nuclear Physics, Homi Bhabha National Institute, Kolkata, India\\
$^{101}$ School of Physics and Astronomy, University of Birmingham, Birmingham, United Kingdom\\
$^{102}$ Secci\'{o}n F\'{\i}sica, Departamento de Ciencias, Pontificia Universidad Cat\'{o}lica del Per\'{u}, Lima, Peru\\
$^{103}$ Stefan Meyer Institut f\"{u}r Subatomare Physik (SMI), Vienna, Austria\\
$^{104}$ SUBATECH, IMT Atlantique, Nantes Universit\'{e}, CNRS-IN2P3, Nantes, France\\
$^{105}$ Sungkyunkwan University, Suwon City, Republic of Korea\\
$^{106}$ Suranaree University of Technology, Nakhon Ratchasima, Thailand\\
$^{107}$ Technical University of Ko\v{s}ice, Ko\v{s}ice, Slovak Republic\\
$^{108}$ The Henryk Niewodniczanski Institute of Nuclear Physics, Polish Academy of Sciences, Cracow, Poland\\
$^{109}$ The University of Texas at Austin, Austin, Texas, United States\\
$^{110}$ Universidad Aut\'{o}noma de Sinaloa, Culiac\'{a}n, Mexico\\
$^{111}$ Universidade de S\~{a}o Paulo (USP), S\~{a}o Paulo, Brazil\\
$^{112}$ Universidade Estadual de Campinas (UNICAMP), Campinas, Brazil\\
$^{113}$ Universidade Federal do ABC, Santo Andre, Brazil\\
$^{114}$ Universitatea Nationala de Stiinta si Tehnologie Politehnica Bucuresti, Bucharest, Romania\\
$^{115}$ University of Cape Town, Cape Town, South Africa\\
$^{116}$ University of Derby, Derby, United Kingdom\\
$^{117}$ University of Houston, Houston, Texas, United States\\
$^{118}$ University of Jyv\"{a}skyl\"{a}, Jyv\"{a}skyl\"{a}, Finland\\
$^{119}$ University of Kansas, Lawrence, Kansas, United States\\
$^{120}$ University of Liverpool, Liverpool, United Kingdom\\
$^{121}$ University of Science and Technology of China, Hefei, China\\
$^{122}$ University of South-Eastern Norway, Kongsberg, Norway\\
$^{123}$ University of Tennessee, Knoxville, Tennessee, United States\\
$^{124}$ University of the Witwatersrand, Johannesburg, South Africa\\
$^{125}$ University of Tokyo, Tokyo, Japan\\
$^{126}$ University of Tsukuba, Tsukuba, Japan\\
$^{127}$ Universit\"{a}t M\"{u}nster, Institut f\"{u}r Kernphysik, M\"{u}nster, Germany\\
$^{128}$ Universit\'{e} Clermont Auvergne, CNRS/IN2P3, LPC, Clermont-Ferrand, France\\
$^{129}$ Universit\'{e} de Lyon, CNRS/IN2P3, Institut de Physique des 2 Infinis de Lyon, Lyon, France\\
$^{130}$ Universit\'{e} de Strasbourg, CNRS, IPHC UMR 7178, F-67000 Strasbourg, France, Strasbourg, France\\
$^{131}$ Universit\'{e} Paris-Saclay, Centre d'Etudes de Saclay (CEA), IRFU, D\'{e}partment de Physique Nucl\'{e}aire (DPhN), Saclay, France\\
$^{132}$ Universit\'{e}  Paris-Saclay, CNRS/IN2P3, IJCLab, Orsay, France\\
$^{133}$ Universit\`{a} degli Studi di Foggia, Foggia, Italy\\
$^{134}$ Universit\`{a} del Piemonte Orientale, Vercelli, Italy\\
$^{135}$ Universit\`{a} di Brescia, Brescia, Italy\\
$^{136}$ Variable Energy Cyclotron Centre, Homi Bhabha National Institute, Kolkata, India\\
$^{137}$ Warsaw University of Technology, Warsaw, Poland\\
$^{138}$ Wayne State University, Detroit, Michigan, United States\\
$^{139}$ Yale University, New Haven, Connecticut, United States\\
$^{140}$ Yonsei University, Seoul, Republic of Korea\\
$^{141}$  Zentrum  f\"{u}r Technologie und Transfer (ZTT), Worms, Germany\\
$^{142}$ Affiliated with an institute covered by a cooperation agreement with CERN\\
$^{143}$ Affiliated with an international laboratory covered by a cooperation agreement with CERN.\\

\end{flushleft} 

\end{document}